\documentclass{aastex7}
\usepackage[nointegrals]{wasysym}
\usepackage{chemformula}
\usepackage[version=4]{mhchem}

\begin{document}

\title{Oxygen Isotope Constraints on the Importance of Photochemical Processing in Protoplanetary Disks}

\author[orcid=00000-0002-0093-065X,sname='Fred Ciesla']{Fred J. Ciesla}
\affiliation{Department of the Geophysical Sciences, University of Chicago, 5734 S. Ellis Avenue, Chicago, IL 60637, USA}
\email[show]{fciesla@uchicago.edu}  

\author[orcid=0000-0002-5954-6302, sname='Van Clepper']{Eric Van Clepper} 
\email{ericvc@uchicago.edu}
\affiliation{Department of the Geophysical Sciences, University of Chicago, 5734 S. Ellis Avenue, Chicago, IL 60637, USA}

\author[0000-0002-8716-0482]{Jennifer Bergner}
\affiliation{UC Berkeley Department of Chemistry, Berkeley, CA 94720, USA}
\email{jbergne@berkeley.edu}

\author[0000-0003-4179-6394]{Edwin Bergin}
\email{ebergin@umich.edu}
\affiliation{Department of Astronomy, University of Michigan, 323 West Hall, 1085 S. University Avenue, Ann Arbor, MI 48109, USA}

\begin{abstract}

Observations have revealed evidence of photochemical processing in protoplanetary disks.  This processing occurs in the photon dominated layer, the optically thin regions of the disk high above the disk midplane.  It remains unclear, however, how much this photochemical processing impacts the compositions of the planets and their building blocks within the disk.  Here we use the oxygen isotopic compositions of Solar System solids, which has been attributed to photochemistry in the solar nebula, to quantitatively evaluate whether this processing could have produced the conditions needed to provide the diversity of compositions seen in the Solar System.  We do this by modeling the chemical evolution while fine dust grows into the building blocks of the planets.  We find that the oxygen isotopic evolution cannot be attributed to processing in the solar nebula and must instead be inherited from the parent molecular cloud.  Further, our results indicate that the observed photochemical processing in protoplanetary disks does not significantly impact the compositions of planets that form within.  

\end{abstract}


\section{Introduction} 

Young stars emit copious amounts of ultraviolet (UV) photons, irradiating their surrounding protoplanetary disks, those of their neighbors, and the molecular clouds within which they are born. These photons are energetic enough that they can  drive disequilibrium chemical reactions. 
Photochemical processing is known to occur in molecular clouds as a result of these photons, with the UV influencing the composition of small hydrocarbons \citep{Cuadrado2015}, which are building blocks for larger molecules; modifying the abundance of water vapor and ice \citep{Hollenbach2009}; and contributing to the overall chemical complexity \citep{Garrod2008}.  Evidence for photochemical processing in protoplanetary disks has also been detected  by the Atacama Large Millimeter Array (ALMA) and the James Webb Space Telescope in the form of transient or reactive ions and radicals \citep{oberg2021,bergner2021, Berne2023}.   While similar reactions may occur in disks as occurs in the clouds, a key difference in this processing is how much material may be affected: molecular clouds are vast, low density objects through which photons can easily penetrate, affecting large amounts of materials.  Protoplanetary disks, however, are much more compact, making them smaller targets, and have higher densities, limiting the photochemistry to the
so-called \emph{photon-dominated layer} at their very upper layers.  Planets and their building blocks, however, form much deeper inside these dust-rich, optically thick disks, around the disk midplanes, where gas and dust densities are highest, facilitating growth.  A key question that remains in our understanding of planetary compositions, then, is: To what extent does the photochemical evolution taking place in the upper regions of a disk impact the compositions of the planetary bodies that eventually form?

Insights into this issue may be gleaned from our own Solar System. Photochemical contributions to the compositions of planets, asteroids, and comets have long been hypothesized, including the formation of organic molecules within simple ices \citep{throop2011,cieslasandford2012}, the production of amorphous water ice and trapping of noble gases within \citep{ciesla2014,mongadesch2014}, and isotopic fractionation of volatile elements \citep{clayton2002,yurimoto2004,young2007,lee2008, mckeegan2011,marty11}. However, while a photochemical origins has been hypothesized, it remains uncertain whether these features were created within the solar nebula via processing of the type seen by ALMA and JWST, or if, instead, they are records of the chemical evolution that occurred within the parent molecular cloud.

Identifying where the photochemical products found in our Solar System were created thus provides an important opportunity for understanding the role that the observed photochemistry in disks has in shaping the planets that form around other stars.  If the photochemical processing occurred in our own protoplanetary disk, then this would suggest the observations of other disks would be providing important insights into how planetary materials were manufactured and altered in those systems.  Alternatively, if irradiation of the solar nebula was shown to be incapable of yielding what we see in the Solar System today, it would indicate that materials  from the molecular cloud were preserved during collapse and play a significant role in setting the compositions of planetary materials \citep{Bergin2024}, and, further, that the observed processing in disks plays a minimal role in shaping planetary properties.

The oxygen isotopic variations seen across Solar System solids provides an opportunity to do exactly that. Laboratory analyses of extraterrestrial materials, such as meteorites and returned samples from small bodies,  provide a quantitative record of the extent, timing, and range of isotopic variations seen across the Solar System.  In the case of oxygen, it has been found that the abundances of the heavy isotopes ($^{17,18}$O) relative to the more common $^{16}$O vary across the planets, asteroids, and comets, with each body seemingly having a characteristic oxygen isotopic composition \citep[e.g.][]{clayton93}.  Importantly, these isotopic variations are mass-independent, indicating that they cannot be explained by kinetic processes, and instead reflect different isotopic abundances in the reservoir of material from which these bodies formed.  The identification of mass-independent variations in oxygen isotopes in Solar System materials were first made five decades ago \citep{clayton1973} and the cause for these variations had been one of the long-outstanding issues in the field of cosmochemistry.  This uncertainty was the primary motivating factor for  the Genesis mission, which sampled the solar wind and returned it to Earth, enabling us to infer the isotopic composition of the Sun \citep{burnett2003}. Analysis of the returned samples showed that the Sun exhibited a much lower $^{17}$O/$^{16}$O and $^{18}$O/$^{16}$O ratios than seen in planetary materials \citep{mckeegan2011}.  This implies that the reservoir of material from which the planets formed was isotopically distinct from the Sun itself, despite both being sourced from the same molecular cloud.  Besides nitrogen, no other element varies across planetary materials to the levels seen in oxygen \citep{clayton93}; most variations in the isotopic ratios among other elements are orders of magnitude smaller.

The enrichments in the heavy isotopes of oxygen observed in planetary materials is largely believed to arise due to isotopically-selective photodissociation of CO during the early stages of planet formation \citep{clayton2002,yurimoto2004,lyons_young_2005,young2007,lee2008, mckeegan2011}.  That is, the different isotopologues of CO require different wavelengths of light to photodissociate.  The greater abundance of C$^{16}$O  in a solar composition gas means that the wavelengths that photodissociate it are readily absorbed when incident on a disk or cloud, while the lower abundances of C$^{17}$O and C$^{18}$O result in the the corresponding dissociative photons penetrating to greater depths. Thus these heavier isotopologues are photodissociated over a greater spatial extent, creating locations where $^{17}$O and $^{18}$O are liberated while $^{16}$O remains bound. The free oxygen can then combine with the abundant hydrogen present to form H$_{2}$O, which is less volatile than CO and would be incorporated into the ice mantles of dust grains. This water would then combine with rocky minerals throughout the Solar System to produce the observed isotopic ratios seen; this combination can occur either through vaporization and recondensation or through gas-solid exchange between water and the primarily amorphous dust that would originally be present in the disk \citep[e.g.][]{yamamoto2018,yamamoto2020,yamamoto2024}.   In fact, evidence that water in the solar nebula was enriched in the heavy isotopes of oxygen has been identified in meteorites \citep{sakamoto2007,marrocchi2018}. 


Previous studies have indeed shown that a reservoir of water enriched in the heavy isotopes of oxygen can be produced in the solar nebula through the photochemical processing described above \citep{lyons_young_2005,young2007}.  Specifically, in the outer nebula, where temperatures are low enough for water to freeze-out into solids while the CO remains in the gas phase, water reservoirs with oxygen isotope ratios consistent with those needed to explain the shift from solar to currently observed values can be formed over hundreds of thousands to millions of years, if the right physical conditions are present.  However, like many astrochemical models, these efforts have not accounted for the changing conditions expected in a protoplanetary disk over these same timescales.  That is, typical models assume unchanging conditions within a protoplanetary disk, whereas the process of planet formation requires the disk to have been dynamic, with dust grains and gas molecules being stirred by turbulence, while planetary building blocks form through the resulting collisions of dust grains.  \citet{vanclepper2022} showed that these processes can drive important feedbacks that impact the chemical evolution of a protoplanetary disk.  For example, dust growth reduces the overall opacity of a protoplanetary disk, allowing UV photons to penetrate deeper into a disk and drive chemical reactions beyond just the very outer surface of the disk.  Turbulent mixing will then carry materials from the deep interiors of the disk to the exposed surface regions, and back again; in addition, when combined with dust growth, this mixing can also lead to changes in the abundances of molecules in the surface of the disk as they may freeze-out and reside in larger dust aggregates that remain around the disk midplane \citep[e.g.][]{krijt2016b}.  All of these would impact the expected chemical evolution that would occur in the disk.

Here, we investigate the production of isotopically heavy water in a dynamic solar nebula to determine the timing and magnitude by which that reservoir would have been created and compare it to the meteorite record.  In doing so, we build on previous studies, investigating for the first time how the coupled effects of turbulent diffusion, coagulation of fine dust into larger aggregates, and the astrochemical processing of gas and ice mantles impacts the isotopic evolution that would occur.  Further, it is the larger aggregates, referred to here as ``pebbles,'' that form from the fine dust grains that serve as the feedstock of planetesimals and planets, and thus it is their chemical and isotopic compositions that are reflected in the objects we observe today. 

In the next section, we describe the model used in our investigation.  We then present our model results, focusing on how dust growth and pebble formation in a diffusive environment impacts the oxygen isotopic evolution expected when compared to previous studies.  We end with a discussion of the implications this work has both for the oxygen isotopic evolution of materials in our Solar System as well as the importance of photochemical processing in protoplanetary disks in shaping the properties of the planets they yield.


\section{Methods} \label{sec:style}

We track the oxygen isotopic evolution of CO and H$_{2}$O as a function of height in a 1D slice of a protoplanetary disk using our Chemistry ANd DYnamics ({\tt CANDY}) model which treats diffusion, dust growth, and chemical evolution as concurrent processes \citep{vanclepper2022}.  We focus on the outer regions of a standard protoplanetary disk \citep{krijt2018}, where CO remains in the gas phase and is subjected to photodissociation while the resulting water is incorporated into the ice mantles of the fine dust present. This dust then aggregates into pebbles, which could then be transported and incorporated into planetary bodies.

While we explore an extensive range of plausible protoplanetary disk conditions, we first focus on a fiducial case to understand how the isotopic evolution is impacted by the suite of processes considered here.   For this, we assume a location of $r$=30 au from the star \citep[similar to ][]{lyons_young_2005}.   The diffusivity of the disk is given by $D = \alpha c_{s} H$, where $c_{s}$ is the local speed of sound, $H$ is the scale height of the gas, and $\alpha$ is a parameter to characterize the level of turbulent mixing; we adopt $\alpha$=10$^{-3}$ for our fiducial case, which is consistent with the upper limits on mixing revealed by ALMA observations \citep{flaherty2018}.
 
Dust is initially present in the disk as  0.1 $\mu$m grains that are well-mixed with respect to the gas with a solid-to-gas mass ratio of 0.01. Over the million-year lifetimes of protoplanetary disks, this dust undergoes significant dynamical and collisional evolution, with fine-dust aggregating into pebbles that decouple from the gas and concentrate around the disk midplane \citep{drazkowska2023}.  To track the formation of these pebbles and the resulting changes in the abundance of fine-dust remaining in the gas, {\tt CANDY} defines a growth timescale which scales with the local dust-to-gas mass ratio, $\epsilon$=$\rho_{d}$/$\rho_{g}$, and orbital frequency, $\Omega$, \citep{birnstiel2012}:
\begin{equation}
    \tau_{g} = a \left( \epsilon \Omega \right)^{-1}
\end{equation}
where $a$ is a parameter describing the efficiency of growth via collisions \citep{krijt2018}; we take $a$=1 for our fiducial run.  Dust growth is assumed to proceed in two phases: (1) an initial ``priming'' period over which dust grains begin to collide with one another but the early-formed aggregates remain small and well-coupled to the gas, and (2) a ``pebble creation'' phase where the larger aggregates form and no longer communicate with the other dust or gas in the disk as they reach either the bouncing or radial drift barriers \citep{birnstiel2012}. The duration of the priming phase is taken to be equal to the growth timescale as determined with the initial dust-to-gas ratio of 0.01;
afterwards the fine dust abundance decreases exponentially as that mass is converted into pebbles.  Pebbles that form at a given time have the same composition as found in the dust at that time throughout the column.\footnote{The exact sizes of pebbles produced in a disk depends on the details of dust growth and whether growth is limited by bouncing, fragmentation, or radial drift \citep{birnstiel2012}, but their sizes are unimportant to the modeling done here.  The only assumption made about pebbles is that once they form, they no longer chemically interact with the dust and gas present in the column; that is, we assume that the surface area available for gas-solid interactions is always dominated by the small dust present.}

The chemical evolution of the disk is modeled using a modified version of the {\tt ASTROCHEM} solver by \citet{maret_bergin2015} as described in \citet{vanclepper2022}.  
The chemical network used here involves a collection of gas-phase reactions, gas-solid exchanges, hydrogenation reactions on grain surfaces, and photochemical reactions.  For solids, only the fine dust is considered in the chemical evolution of the disk as this would dominate the available grain surface area on which reactions could take place. Ice mantles on these grains are assumed to be mixed, such that the molecules available to interact with the gas are present in a proportion equal to that throughout bulk mantle (i.e. there is no compositional layering of the ice mantle).  Isotopic-specific self-shielding of CO in the gas is treated as described in \cite{visser2009}. 

To avoid any uncertainties and complexities associated with a full chemical network, we adopt a simplified (reduced) network that focuses on key reactions that lead to transfer of oxygen from gaseous CO to solid water ice, namely the dissociation of CO to C and O, and possible pathways for that O to either reform CO or react to form H$_2$O (Figure 1). The reactions are shown generically for oxygen, but in reality, three clones of each equation are included for $^{16}$O, $^{17}$O, and $^{18}$O, with all reaction rates being identical (accounting for mass dependence when appropriate).  

Water formation occurs on grain surfaces via hydrogenation as in \cite{visser2011}; gas phase formation of water is slow by comparison.  This requires a hydrogen atom to collide with a frozen-out oxygen atom on a grain surface to yield OH on the grain, then another H atom to collide with the frozen-out OH to yield H$_{2}$O.  The efficiency of this process is largely controlled by the residence time of the target O and OH on grain surfaces; shorter residence times decrease the likelihood of a hydrogenation reaction occurring and moving towards the production of water.  The residence times of the species are largely set by the binding energies for the thermal desorption reactions.  Binding energies are generally found experimentally, and the details vary from species to species and on the composition and physical structure of the surface on which the species resides \citep{minissale2022}.  
We adopt high binding energies for O, OH, and H$_{2}$O to simulate very rapid formation of water from liberated oxygen by taking the standard binding energies from \citep{umist2012} for each and increasing them by two orders of magnitude.  This conservative treatment likely overestimates the production rate of water, allowing us to focus on the consequences of the dynamical effects in the disk; as we find water production rates are not the limiting factor in the evolution, this treatment does not alter our conclusions.

\begin{figure}[t]
    \centering
    \includegraphics[scale=0.25]{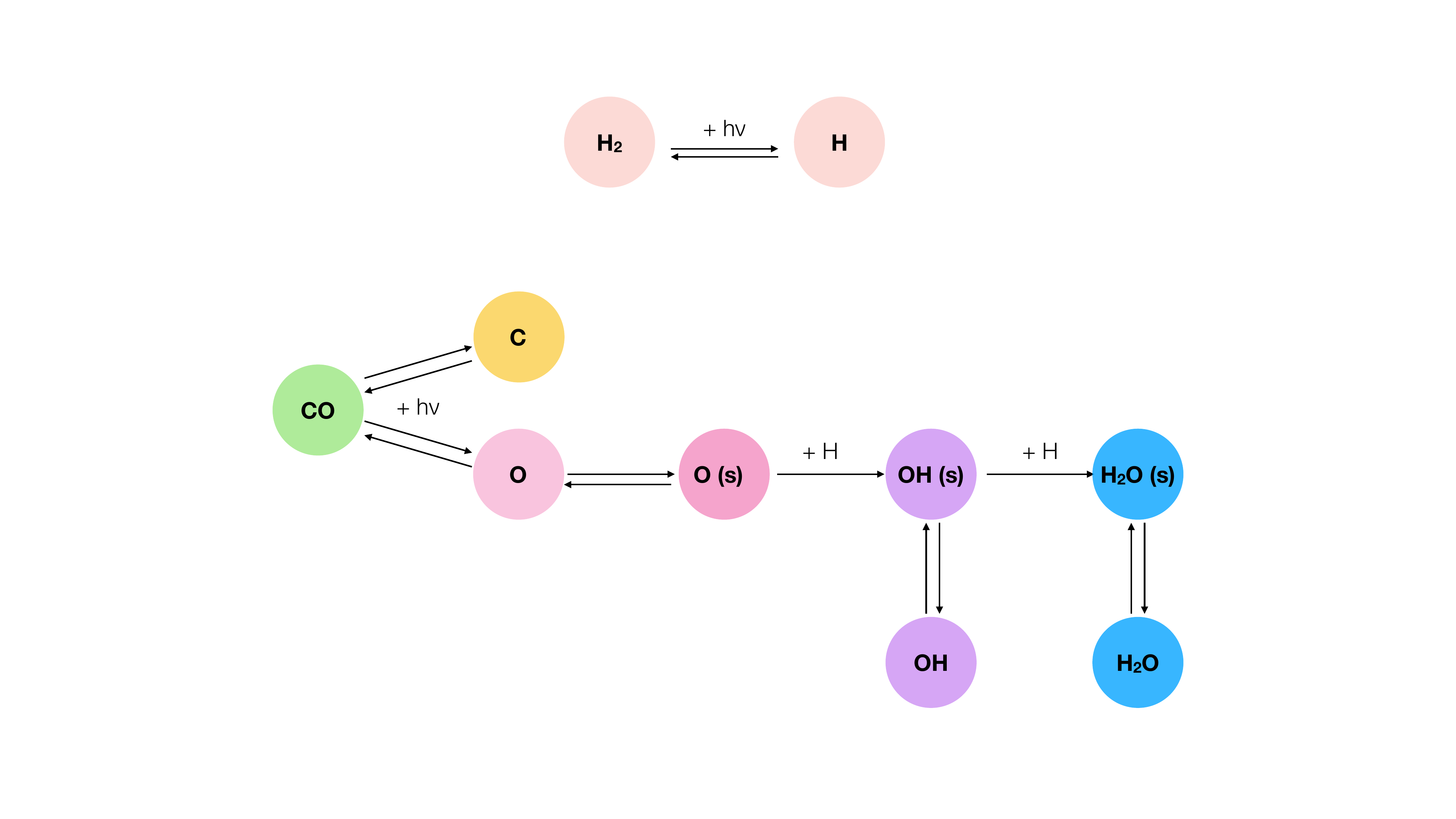}
    \label{fig:network} 
    \caption{Illustration of the reaction network used in this study.  Parameters for determining reaction rates are given in Table 1.}
\end{figure}

\begin{center}
\begin{table}[!h]
\begin{tabular}{ l c c c c c c c c}

Reaction & & $k_{0}$$^{a}$ & $\alpha$$^{b}$ & $\beta$$^{b}$ & $\gamma$$^{b}$ & $E_{b}$(used)$^{c}$ & $E_{b}$(standard)$^{c}$ \\
 \hline
H$_{2}$ + $h \nu$ $\rightarrow$ H + H &  & 5.80 $\times$10$^{-11}$ & - & - & - & - & - \\
CO + $h \nu$ $\rightarrow$ C + O &  & 2.00 $\times$10$^{-10}$ & - & - & - & - & - \\
H + H $\rightarrow$ H$_{2}$ & & - & 3.00$\times$10$^{-18}$ & 0.50 & 0.00 & - & - \\
C + O $\rightarrow$ CO & & - & 4.69$\times$10$^{-19}$ & 1.52 & -50.5 & - & - \\
O $\rightarrow$ O (ice) & & - & 1.00 & 0 & 0 & - & - \\
OH $\rightarrow$ OH (ice) & & - & 1.00 & 0 & 0 & - & - \\
H$_{2}$O $\rightarrow$ H$_{2}$O (ice) & & - & 1.00 & 0 & 0 & - & - \\
O (ice) $\rightarrow$ O & & & - & - & - & 1.75$\times$10$^{5}$ & 1.75$\times$10$^{3}$ \\
OH (ice) $\rightarrow$ OH & & & - & - & - & 2.85$\times$10$^{5}$ & 2.85$\times$10$^{3}$ \\
H$_{2}$O (ice) $\rightarrow$ H$_{2}$O & & & - & - & - & 4.80$\times$10$^{5}$ & 4.80$\times$10$^{3}$ \\
O (ice) + $h\nu$ $\rightarrow$ O & & & 1.00$\times$10$^{-3}$ & - & - & - & - \\
OH (ice) + $h\nu$ $\rightarrow$ OH & & & 1.00$\times$10$^{-3}$ & - & - & - & - \\
H$_{2}$O (ice) + $h\nu$ $\rightarrow$ H$_{2}$O & & & 1.00$\times$10$^{-3}$ & - & - & - & - \\
O(ice) + H $\rightarrow$ OH (ice) & & & 1.00 & 0 & 0 & - & - \\
OH (ice) + H $\rightarrow$ H$_{2}$O (ice) & & & 1.00 & 0 & 0 & - & - \\
 \hline \\
\end{tabular}
\quad \\
\caption{Reactions and reaction rate parameters for the astrochemical network used here.  Note all reactions with oxygen are cloned for $^{17}$O and $^{18}$O with the same reaction rate parameters listed above. All rates follow the parameterization in \cite{umist2012}. $^a$ Unattenuated dissociation rates taken from \cite{balashev2020} for H$_{2}$ and \cite{umist2012} for CO.
$^b$ Reaction rate parameters taken from \cite{umist2012}. $^{c}$ Two values of binding energies are shown for for O, OH, and H$_{2}$O, the value used in our model to allow for efficient formation of water and the standard value from the literature it is based on.  The binding energies for the low efficiency case are taken from \cite{minissale2022} for O and \cite{umist2012} for OH and H$_{2}$O.  The efficient binding energies are found by increasing $E_{b}$ by two orders of magnitude over those cited in the literature. Photodesorption rates are all assumed to have yields given by $\alpha$ and parameterized as described in \cite{maret_bergin2015}. Hydrogenation reactions are calculated as described in \cite{visser2009} and \cite{bosman2018}}
\end{table}
\end{center}

{\tt CANDY} performs the chemical and dynamic (diffusion and dust growth) calculations in separate steps.  That is, the chemical evolution is calculated at each location in the column for a period of time, $t_{chem}$, then the dust and gas are allowed to diffuse, as well as pebbles to grow, for that same period of time in alternating steps over a shorter time periods, $t_{dyn}$.  We use $t_{chem}$=500 years here and alternate between diffusion and growth over a period of $t_{dyn}$= 100 years as we have found that these choices allow the calculations to be performed in a reasonable amount of time while still yielding convergent results with runs where the calculations are alternated on shorter timescales \citep{vanclepper2022}.

We begin all runs with H$_{2}$O and CO having mixing fractions $X_{\mathrm{H_{2}O}}$=1.18$\times$10$^{-4}$ and $X_{\mathrm{CO}}$=10$^{-4}$ \citep{bosman2018,vanclepper2022} where $X_{i}$=$n_{i}$/$n_{H}$ with $n_{H}$ being the total amount of hydrogen atoms present at a given location.  We further assume the same oxygen isotopic composition for the H$_{2}$O, CO, and Silicates as the Sun, $\delta^{17,18}$O = -59 \permil \citep{mckeegan2011}; this is done to focus on whether the observed isotopic ratios can be reached purely as a result of effects within the solar nebula.  This  would not be the case if CO self-shielding was a significant process in the molecular cloud \citep{yurimoto2004,lee2008}, or if the molecular cloud was recently polluted by ejecta from a nearby massive star \citep{krot2010}.

\section{Results}

\subsection{Fiducial Case: Isotopic Evolution in a Dynamic Disk}

Figure 2 shows the progression of the isotopic composition of H$_{2}$O and CO in the disk as a function of height above the disk midplane in the case where dust growth does not occur.  Note, we plot only $\delta^{18}$O values for CO and H$_{2}$O because the $\delta^{17}$O values are indistinguishable from them.  In this \emph{no growth} model, the photon-dominated layer is readily seen in the upper layers of top panel as all CO isotopologues are destroyed above $z \gtrsim 4H$.  Because photons of all wavelengths are present in this region,  all isotopologues of CO are dissociated, yielding new water with the same isotopic composition as the bulk system.  The self-shielding region of CO, where the heavy isotopologues (C$^{18}$O, C$^{17}$O) continue to be photodissociated but C$^{16}$O is not, is just below this, $3.5 H \lesssim z \lesssim 4 H$.   The water produced in this region is enriched in the heavy isotopes of oxygen, resulting in greater $\delta^{18}$O. In contrast, the CO becomes depleted in those isotopes, driving its $\delta^{18}$O to lower and lower values.  Throughout the simulation, this region of the disk continues to be a source of heavy water as this is where the dissociating photons for the heavy isotopologues are able to penetrate to, and the CO is replenished as it diffuses upwards from deeper layers of the disk.   Diffusion also mixes the photochemically-produced heavy water downward, raising the $\delta^{18}$O values of H$_{2}$O and reducing those of CO all the way down to the midplane.  The column-averaged evolution of H$_{2}$O and CO as a function of time is shown in the bottom panel of Figure 2.

\begin{figure}[!h]
    \centering
    \includegraphics[width=5in]{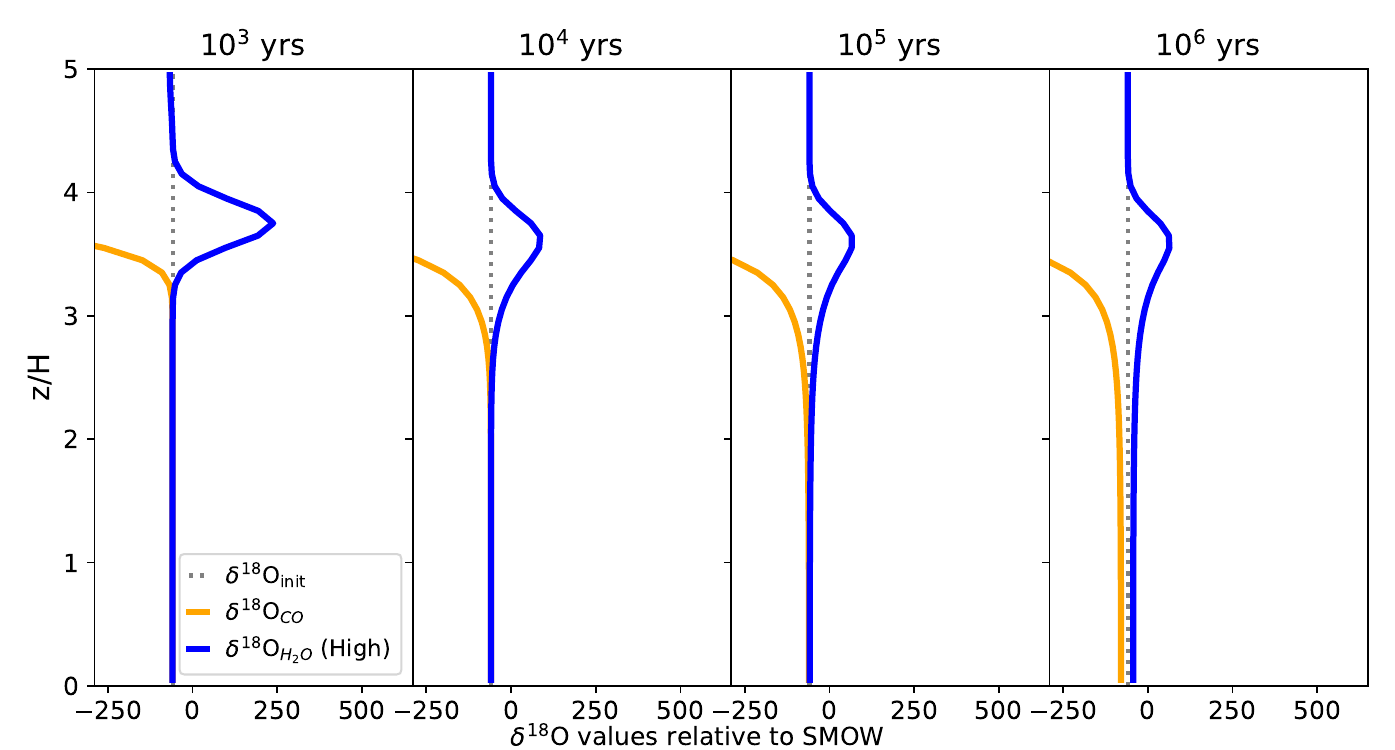} \\
    \includegraphics[width=5in]{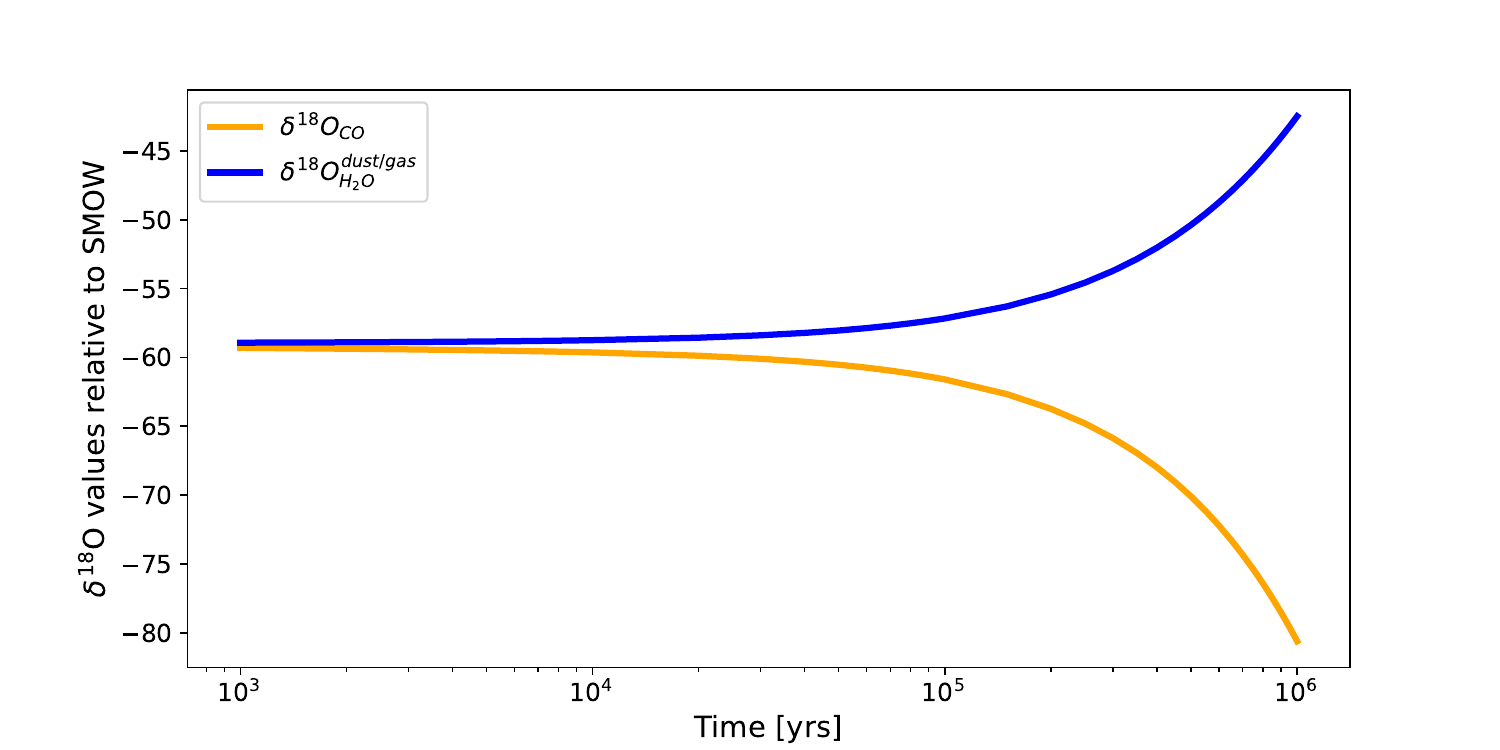} 
    \label{fig:fiducial} 
    \caption{Oxygen isotopic evolution in the \emph{no growth} model described in the text.  Top: Oxygen isotopic composition of H$_{2}$O and CO as a function of height in our model disk at various times throughout the simulation with no dust growth. The $\delta^{17}$O values evolve in essentially identical way to $\delta^{18}$O, so only the latter are shown here.  The vertical dotted line represents the starting isotopic compositions of the two species.  The $\delta^{18}$O of CO goes to -$\infty$ as all CO is destroyed at the very surface of the disk. Bottom: Column averaged isotopic evolution of the H$_{2}$O and CO over time in the model.}
\end{figure}

Figure 3 shows the results for the same fiducial run, but with dust growth allowed to occur. The results after just $\sim$10$^{3}$ years are identical to the \emph{no growth} model because this is the priming period of dust growth; after this time, though, the abundance of dust decreases as pebbles form, resulting in very different evolution of the H$_{2}$O and CO reservoirs than in the \emph{no growth} case.  This occurs because small grains are the main source of UV opacity in the disk, and their decreased abundance allows all photons to penetrate to greater depths \citep{vanclepper2022}.  As a result, the peak in the $\delta^{18}$O composition of the water, that is the region where C$^{17}$O and C$^{18}$O are photodissociated but C$^{16}$O is not, migrates downwards with time.  Notably, the values of the $\delta^{18}$O for water increase significantly in this \emph{growth} case compared to the values seen in the \emph{no growth} case.  This can be attributed, in part, to the greater amounts of CO being photodissociated within the column as a result of the self-shielding region migrating deeper inside of the disk, amplifying the isotopic selective effect.  

\begin{figure}[!t]
    \centering
    \includegraphics[width=5in]{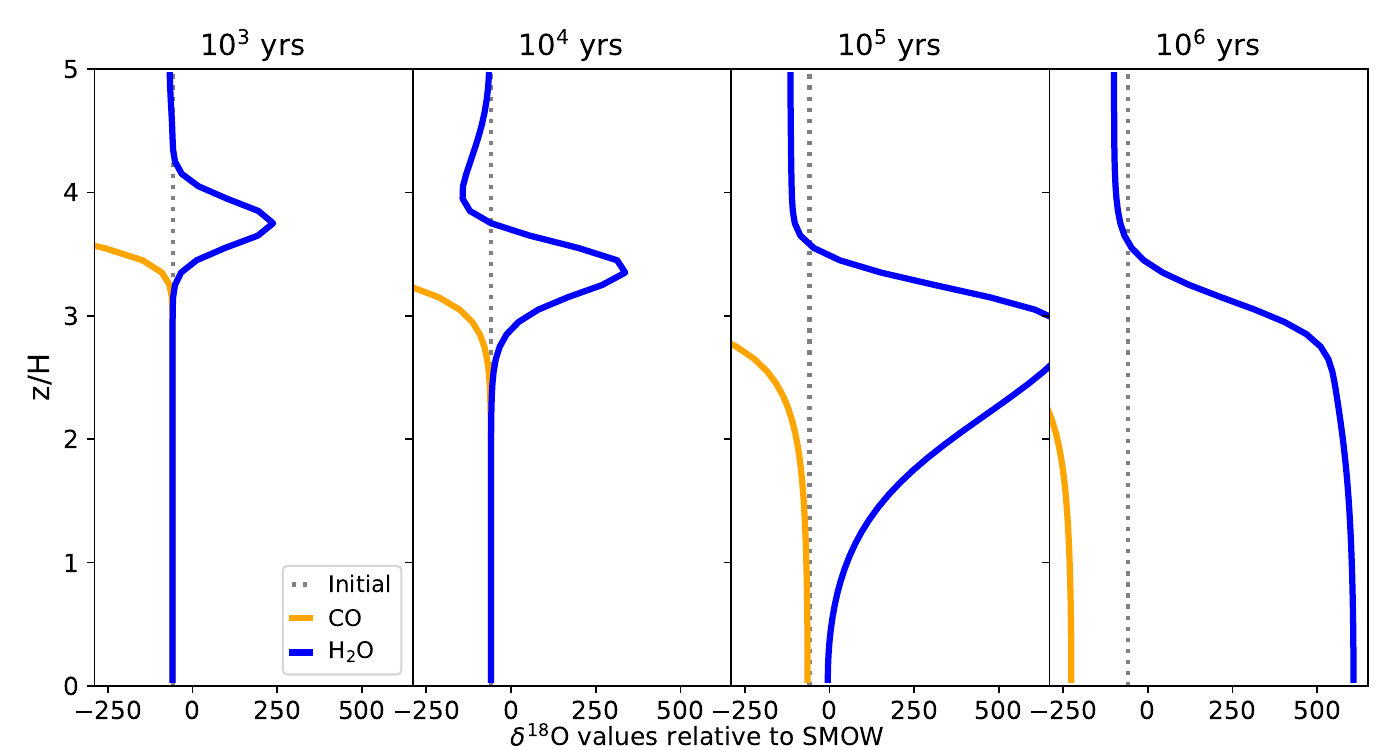}\\
    \includegraphics[width=5in]{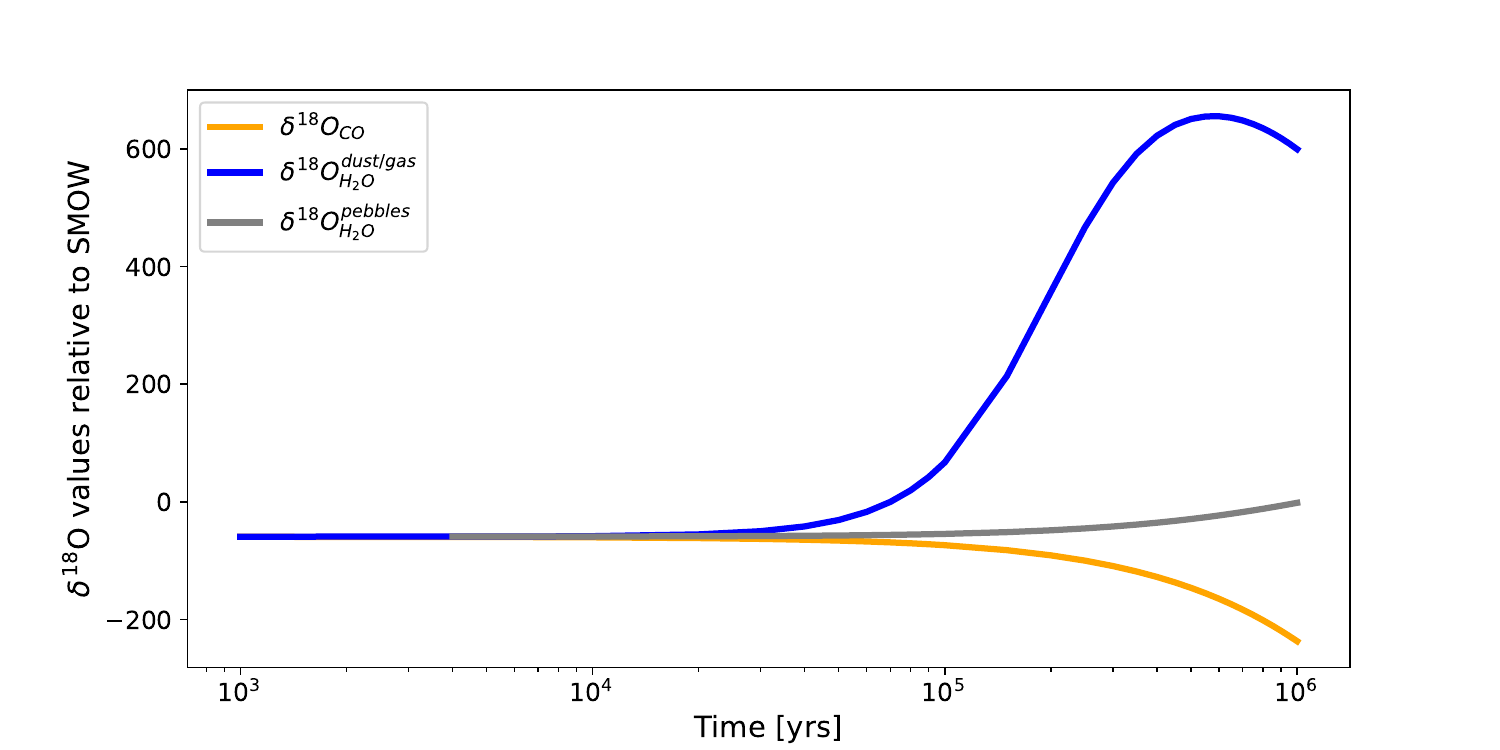} 
    \label{fig:fiducial2} 
    \caption{Same as Figure 2, but for the \emph{dust growth} case.  The grey line in the bottom panel reflects the cumulative isotopic evolution of the pebbles that grow in the disk.}
\end{figure}

However, it is important to note that the values shown in the top panel of Figure 3 reflect the isotopic compositions of CO and H$_{2}$O contained in the chemically-active portion of the disk: the gas phase or in the ice mantles of the fine dust in the column.  As shown in the bottom panel of Figure 3, the pebbles follow a different path.  The isotopic evolution of the water contained in the pebbles lags that seen in the dust and vapor; in this case reaching a maximum of near $\delta^{18}$O$\sim$0\permil in the time considered, while the water remaining in the gas and dust reaches $\delta^{18}$O$\gtrsim$ +600\permil.  It must be remembered that the water contained in the pebbles is a mixture of the isotopically light water that was originally present in the disk and the isotopically heavy water produced as a result of CO self-shielding.  At early times, photoprocessing is limited to the very upper layers of the disk due to the high opacity provided by the dust.  While this results in a reservoir of isotopically-heavy water being created, it is a region that contains only a small fraction of the total column; approximately 0.3\% of the disk mass is contained above $z$=3$H$.  Most of the mass of dust that forms the pebbles comes from deeper within the disk, around the disk midplane. At early times, the water that the pebbles contain is thus dominated by the isotopically-light water that was initially present in the nebula; only a small amount of heavy water was produced.  As a result, the early-formed pebbles will contain water with $\delta^{18}$O=-59 \permil.  

Thus, as pebbles form early, they preferentially remove isotopically-light water from the column.  As time goes on, more and more photochemically-produced heavy water forms, growing in abundance in both relative and absolute terms.  This allows much higher $\delta^{18}$O values to be reached in the column compared to the \emph{no growth} case where the heavy water was constantly mixed with the original, isotopically-light water in the disk.  It is the later formed pebbles that will contain water with high $\delta^{18}$O values.  However, at these later times, the rate of production of pebbles decreases significantly because the dust is depleted, limiting the raw materials from which they would form.  This means that the pebbles that contain significant photochemical products are relatively low in abundance compared to the earlier-formed pebbles.


\begin{figure}[!h]
    \centering
    \includegraphics[width=5in]{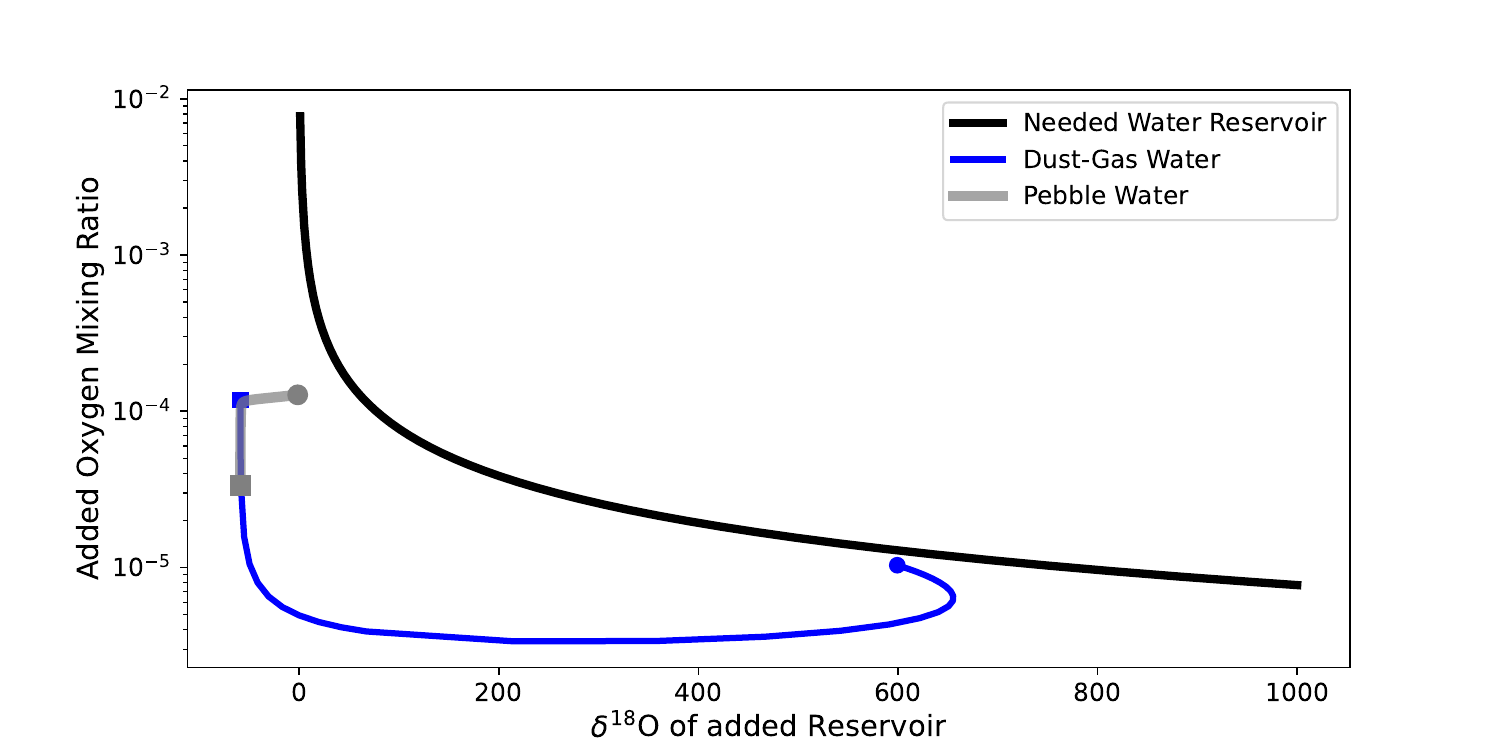}
    \label{fig:needed_comp} 
    \caption{Comparison of the evolving water reservoirs contained in pebbles and mantles of dust grains to the reservoir needed to mix with a initially solar mix of silicates to produce the oxygen isotope ratios observed in Solar System rocks.  The mixing ratio is defined as the ratio of the water to total hydrogen atoms present, $X_{\mathrm{H_{2}O}}$=$n_{\mathrm{H_{2}O}}$/$n_{H}$}.
\end{figure}

Given that photoprocessing leads to the production of isotopically-heavy water,  we next investigate whether the combined isotopic composition and abundance of water in this reservoir is sufficient to drive the silicates and oxides found in meteorites and terrestrial planets from solar-like oxygen isotopic compositions  to the values seen today.  In Figure 4, the black line indicates abundances and isotopic compositions of water needed to be added to a solar mixture of silicates (in abundance and isotopic composition) to produce a net product with $\delta^{18}$O$\approx$0 \permil (the reference value for oxygen on Earth and a value around which most extraterrestrial samples cluster).
These values were found by adopting a canonical oxygen-to-hydrogen ratio for a solar composition gas (O/H$\sim$5.75$\times$10$^{-4}$) and assuming 22.75\% of that oxygen is contained by rocky minerals \citep{lodders2003}.  We then assign the abundances of each isotope of oxygen in these rocks based on the solar ratio as given by \citet{mckeegan2011}.   As expected, for a water reservoir that is high in $\delta^{18}$O, a smaller amount is needed to mix in to yield the inferred shift in isotopic ratios, whereas a greater amount is needed if the water reservoir is not as enriched in the heavy isotopes of oxygen.

To compare the needed water reservoir with that produced within the disk, Figure 4 also shows the evolution of the isotopic and mixing ratio of the dust / gas and pebbles in our model.  The blue line represents the path of the water reservoir carried by the fine dust and gas within the column, beginning at the square at ($\delta^{18}$O,$X_{\mathrm{H_{2}O}}$)=(-59\permil, 1.18$\times$10$^{-4}$), and evolving to the circle over the 10$^{6}$ year period consider here. 
Again, this reservoir reaches very high $\delta^{18}$O values, but in doing so, the abundance of the water drops throughout much of this time.  There is an increase in the water abundance late in the simulation, when the rate of water production from photodissociated CO exceeds the rate at which that water is removed via pebble formation.  We note that eventually the $\delta^{18}$O values of the water begin to decrease from their maximum value. This occurs after a few hundred thousand years when the heavy isotopologues of CO become sufficiently depleted in the disk, resulting in the photodissociated CO to be increasingly dominated by the lightest isotopologue, C$^{16}$O. From the evolution shown, we see that the water reservoir contained in the dust and gas at any given time would be insufficient to combine with solar-composition rocks to yield the observed oxygen isotopic values observed today.

The grey line in Figure 4 represents the evolution of the water reservoir carried by the pebbles, which are what actually would be transported from this region of the disk and made available to rocky bodies in the inner part of the Solar System. Again, the evolution begins with the square representing the first appearance of pebbles around $\sim$3000 years and the circle being where things evolved to after 10$^{6}$ years.   As discussed above, the water reservoir in the pebbles is dominated by the original inventory of water present in the disk; this reservoir moves in a nearly vertical line upward in the plot (roughly constant oxygen isotopic composition) as more pebbles form and minimal amounts of heavy water are incorporated.  The pebble reservoir does eventually incorporate the heavy water produced in the disk, and the bulk isotopic composition moves to higher $\delta^{18}$O values, though not significantly when compared to that remaining in the gas and dust.  Some additional water is created within the disk, providing more than was present at the start of the simulation; it is a small amount, increasing the total water by $\sim$16\% at the end compared to what would be there without any photochemical processing.  As a result, this reservoir is never sufficient to mix with rocks to form the isotopic compositions observed today.

\subsection{Other Disk Conditions and Parameters}

As the conditions within a protoplanetary disk and the details of dust growth remain uncertain, we carried out a number of additional simulations to determine how the outcomes depended on the parameters used in the model.   Here we describe how our results vary with the different parameters used in our model and evaluate the possibility that the water reservoir created in the disk could be responsible for helping to evolve oxides and silicates with solar oxygen isotopic ratios to the values seen today.

\subsubsection{Dust Growth Timescale} Pebbles in our model are assumed to form via collisional aggregation of dust grains, and the details of these collisions and how they lead to growth depends on a number of factors such as the composition of the dust, the porosity of the aggregates, and the level of turbulence in the disk \citep[see discussion in][]{krijt2018}.  To explore the effects of different rates of dust growth, we ran simulations with $a$=1, 10, and 100, with $a$=1 being the fiducial model described above.  The resulting isotopic evolution of the H$_{2}$O and CO reservoirs are shown in Figure 5.

While slower dust growth allows more time for chemical evolution to occur before pebbles are formed, the resulting higher abudances of dust over time limits photoprocessing to the very upper, diffuse regions of the disk.  As this is where densities are very low, this prevents significant amounts of CO from being photodissociated, and thus hinders the formation of a heavy water reservoir.  As a result, the water reservoirs are always insufficient to yield a shift in rocky minerals from solar to the values observed today (Figure 6).  Faster growth would lead to more efficient pebble formation prior to the photochemistry being able to create the isotopically heavy water, meaning the pebbles would largely just lock away the starting composition of the nebula.

 \begin{figure}[!h]
     \includegraphics[width=3.3in]{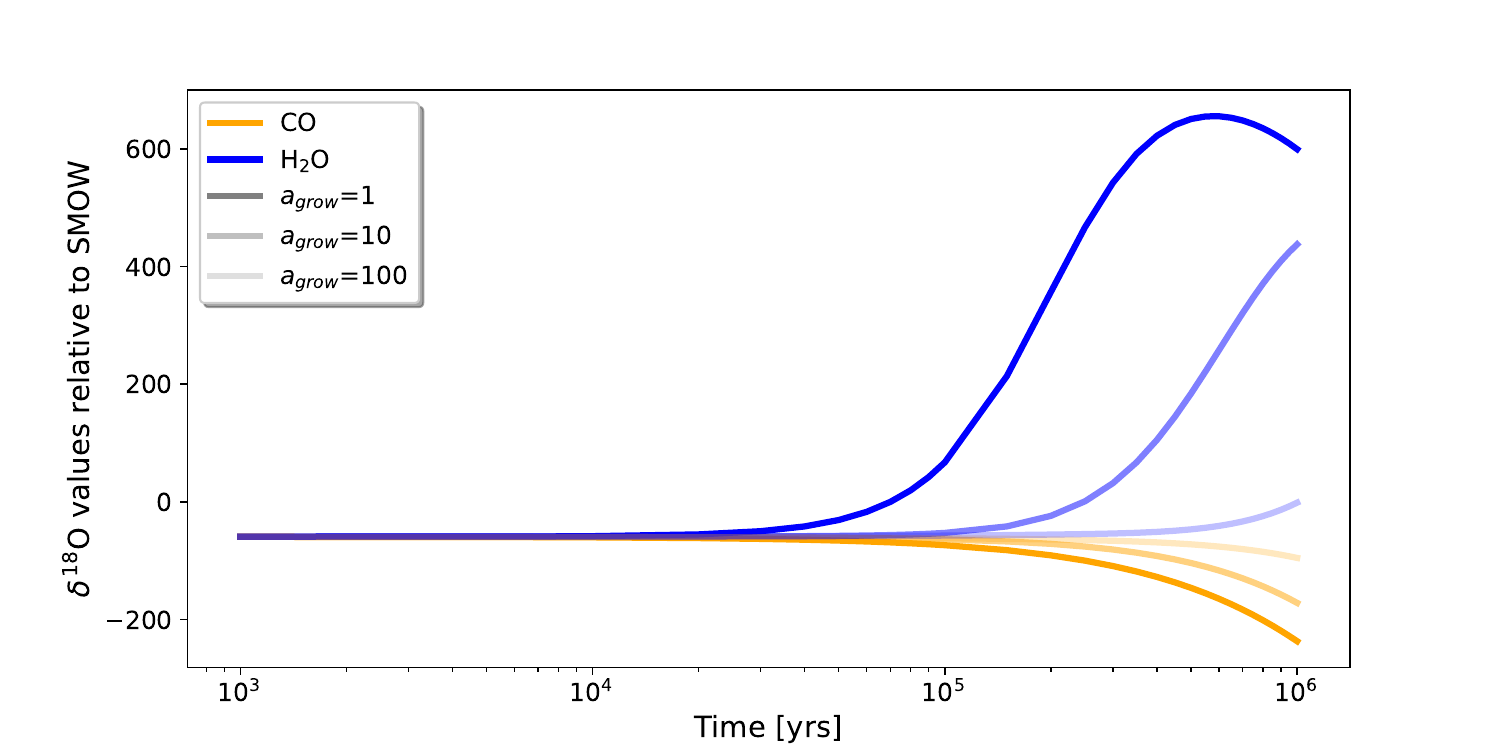} 
    \includegraphics[width=3.3in]{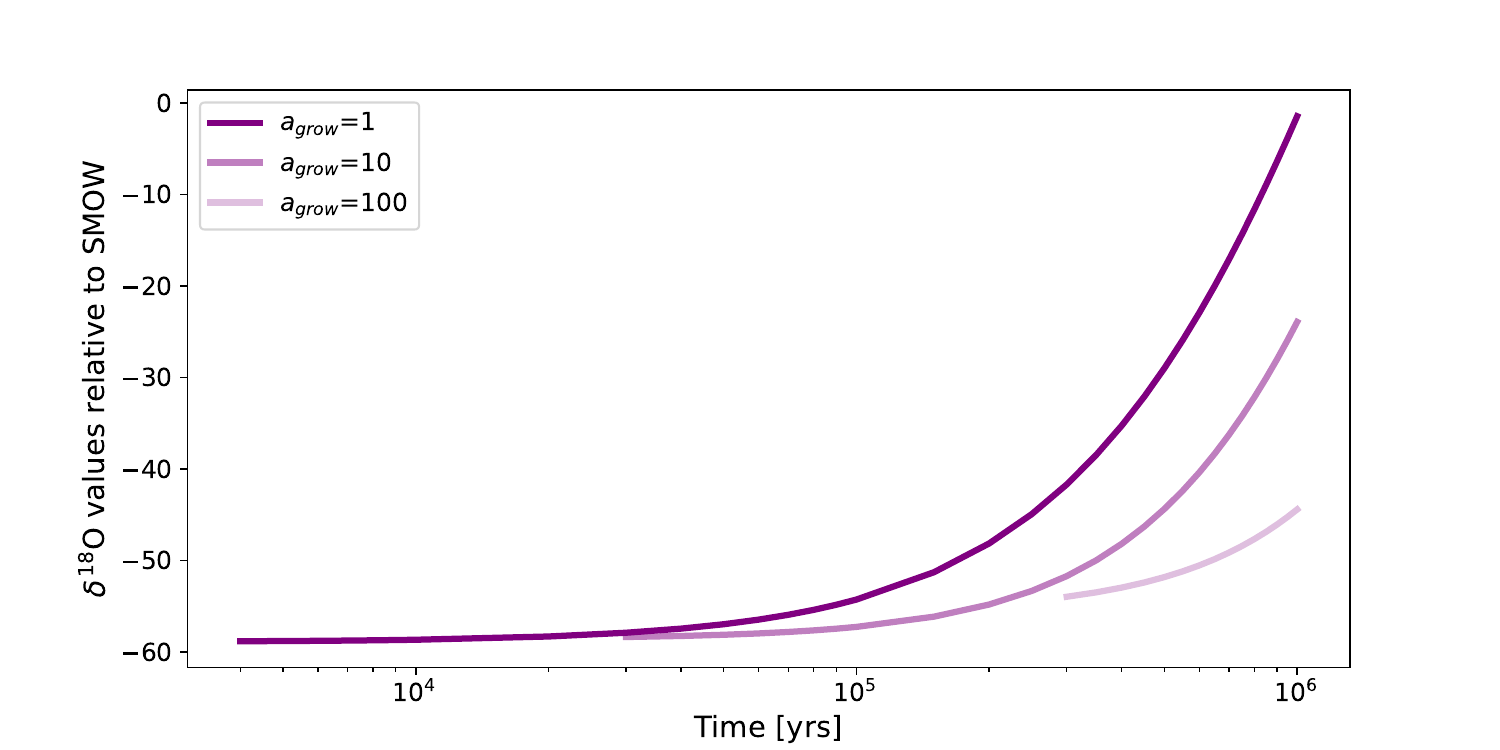} \\

    \label{fig:fig_params} 
    \caption{Isotopic evolution of the CO and H$_{2}$O reservoirs in the disk (left) and pebbles (right) for various rates of dust growth.}
\end{figure}

 \begin{figure}[!h]
\includegraphics[width=2.3in]{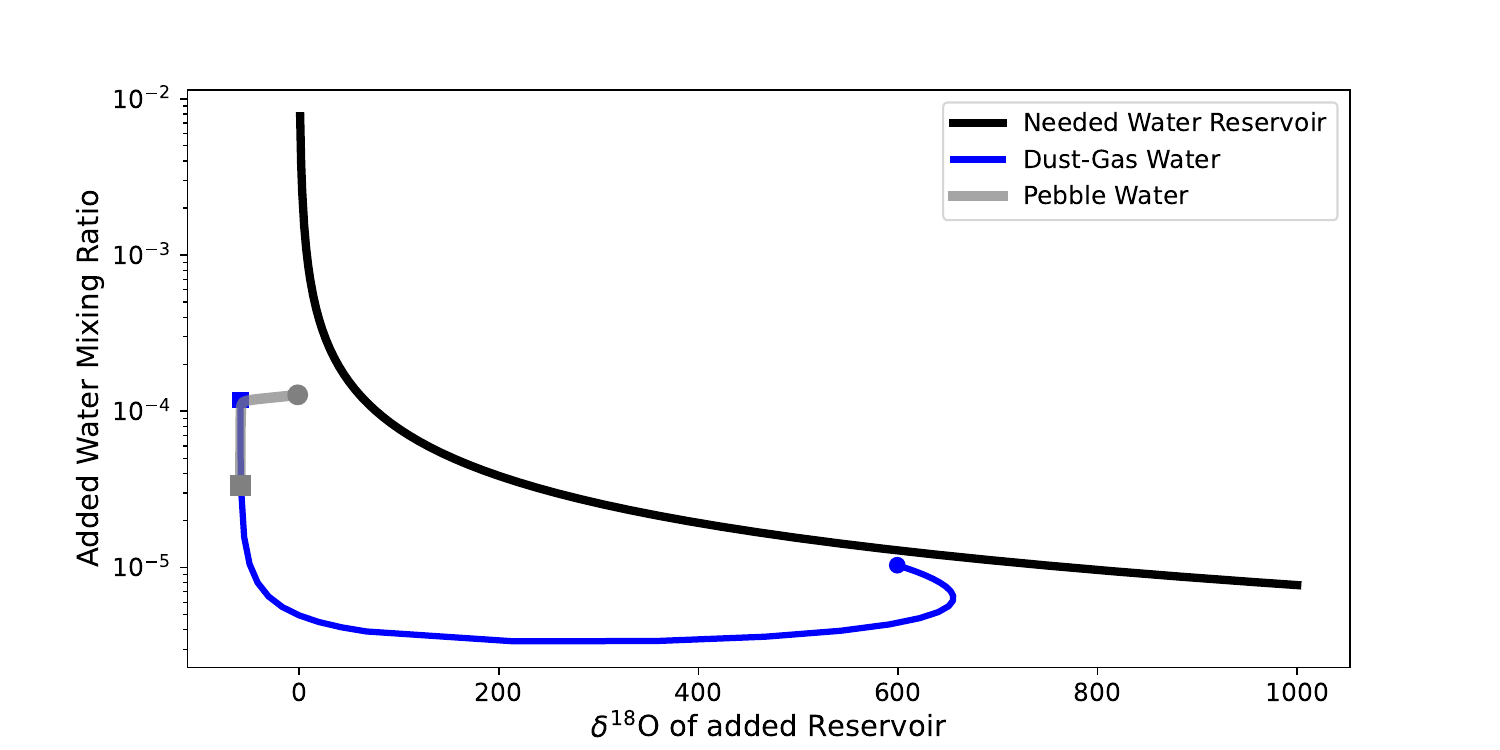}
\includegraphics[width=2.3in]{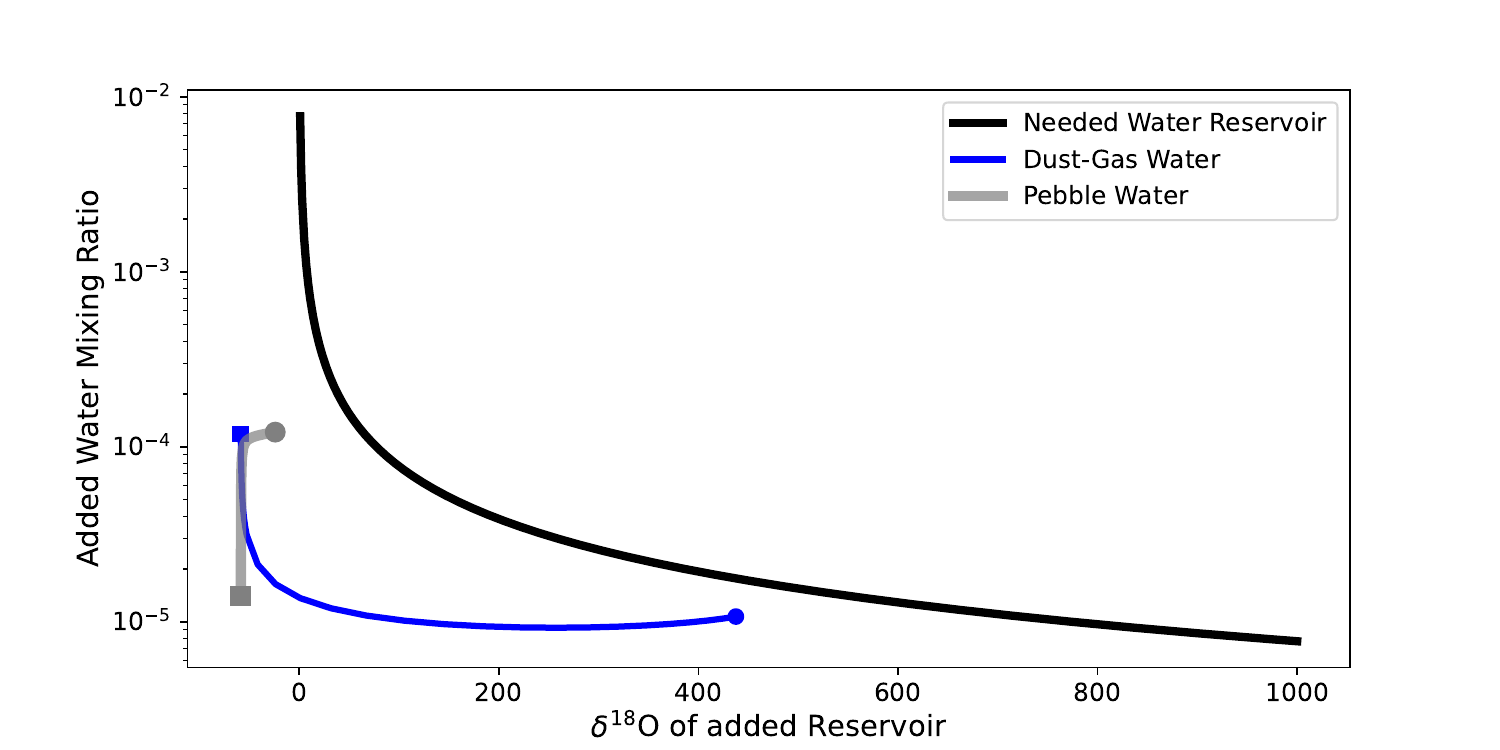}
\includegraphics[width=2.3in]{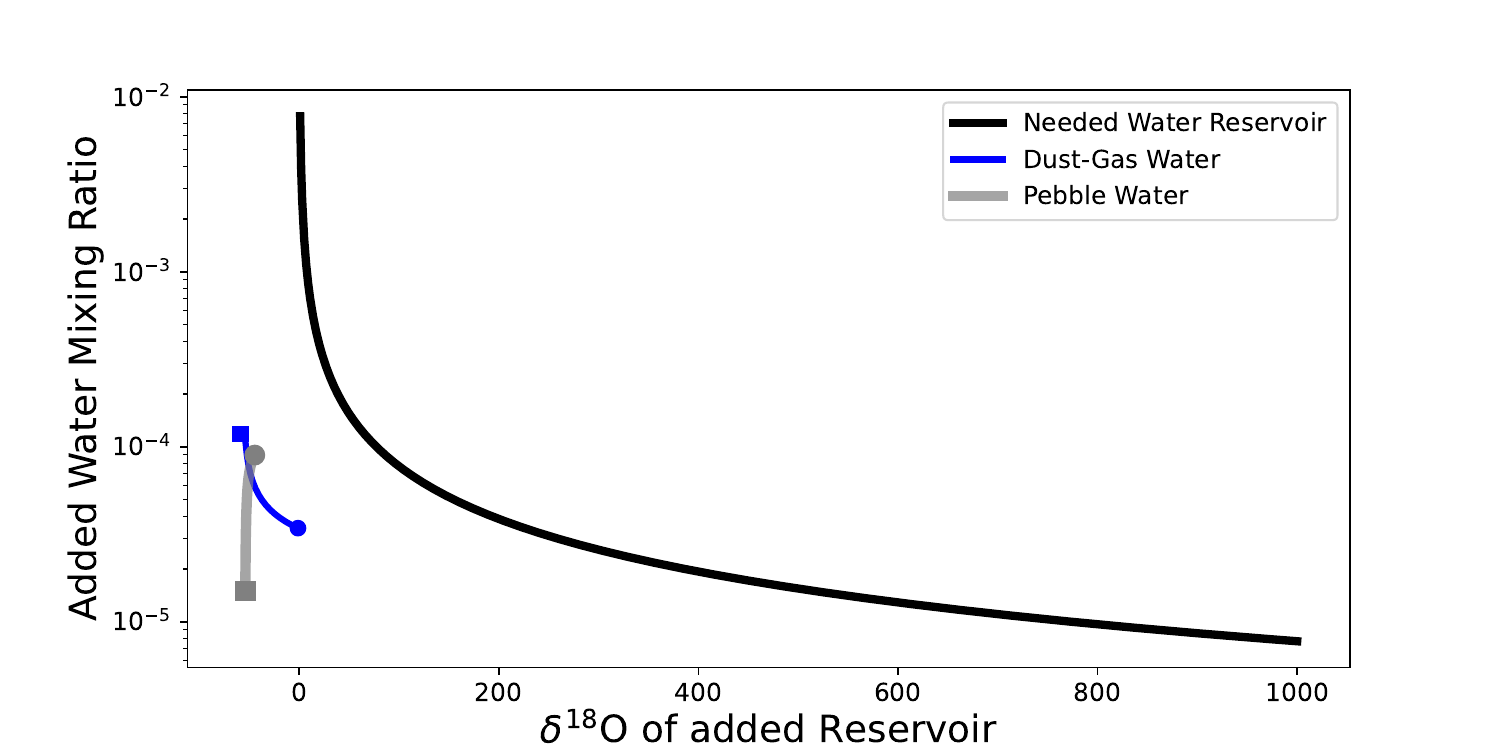} 
    \label{fig:grows} 
    \caption{Comparison of the isotopic evolution of the water reservoirs in the various dust growth rate cases compared to that needed to shift the silicates and oxides of solar composition to terrestrial values, as shown in Figure 4.  The runs are $a$=1 (our fiducial run; left), $a$=10 (middle) and $a$=100 (right).}
\end{figure}

\subsubsection{UV Radiation Flux} Figure 7 shows the isotopic evolution expected with a vertical UV radiation field of $\chi$=50, 500, and 5000 in Draine units \cite{draine78}, with $\chi$=50 being the value used in our fiducial case. More intense radiation fields provide more photons to dissociate CO, which results in the greater water mixing ratios produced in Figure 8.  However, while the water is enriched in heavy isotopes as a result of self-shielding in each case, the more intense radiation fields yield lower $\delta^{18}$O reservoirs (also found by \citet{lyons_young_2005}).  This occurs because the greater photon flux allows for more efficient destruction of C$^{16}$O.

 \begin{figure}[!h]
        \includegraphics[width=3.3in]{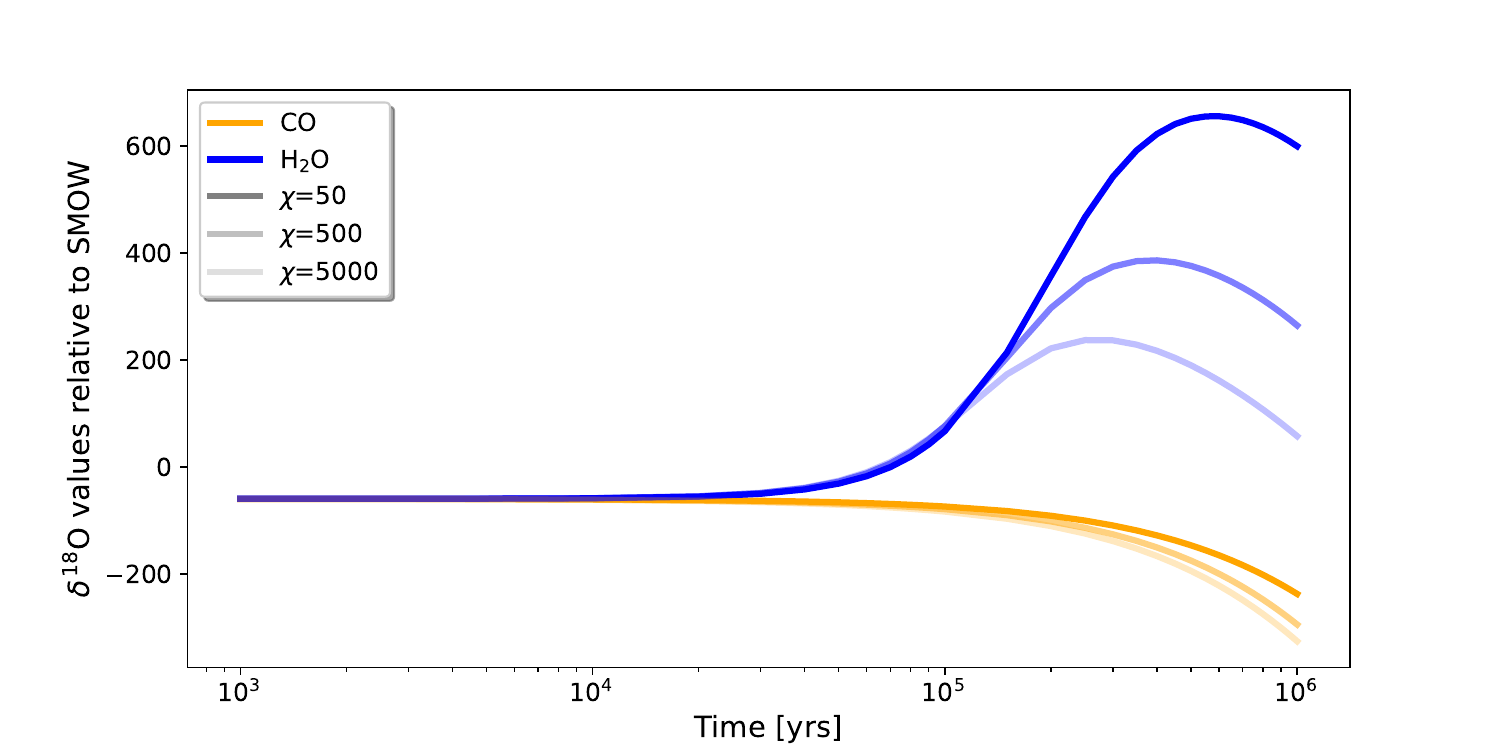} 
    \includegraphics[width=3.3in]{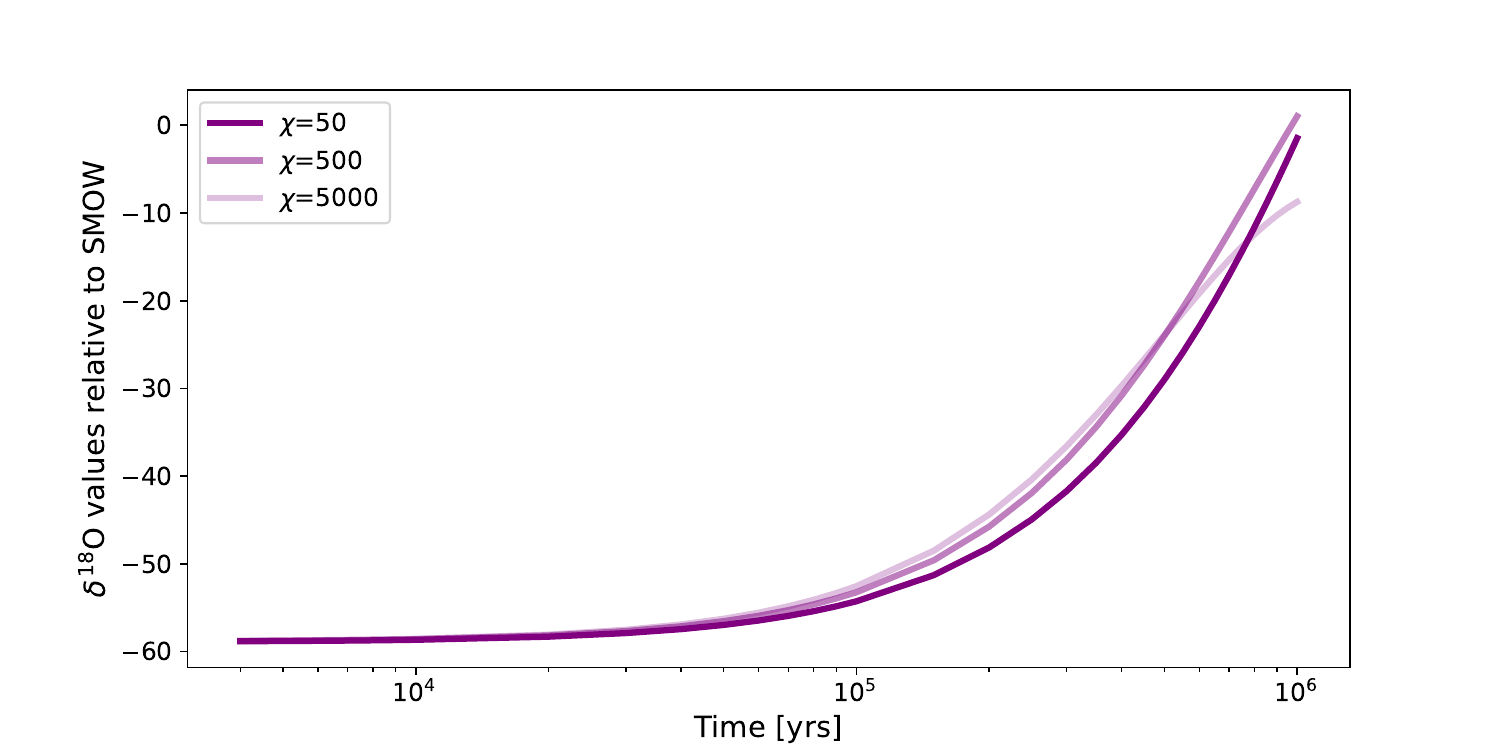} \\
    \label{fig:fig_uv} 
    \caption{Isotopic evolution of the CO and H$_{2}$O reservoirs in the disk (left) and pebbles (right) for various UV radiation field intensities.}
\end{figure}

 \begin{figure}[!h]
\includegraphics[width=2.3in]{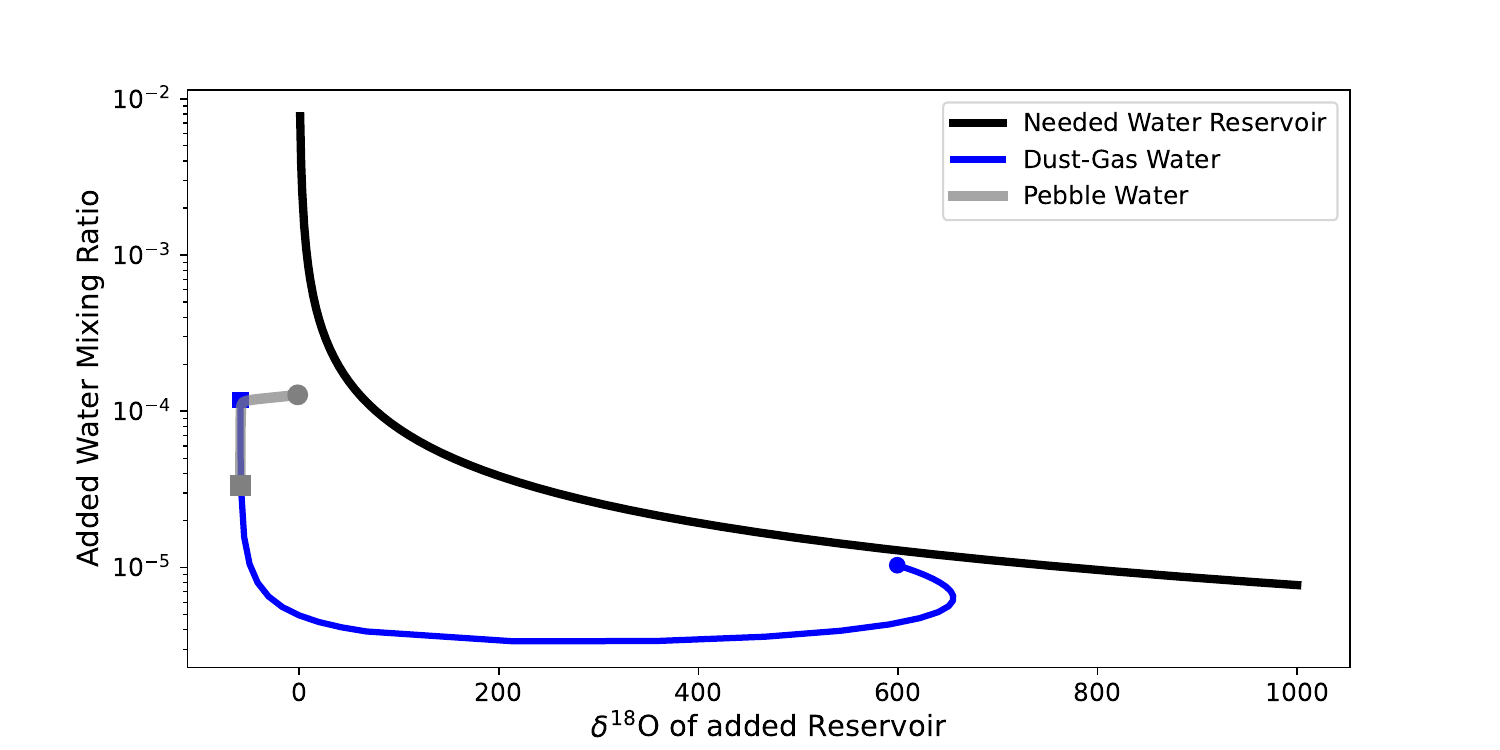}
\includegraphics[width=2.3in]{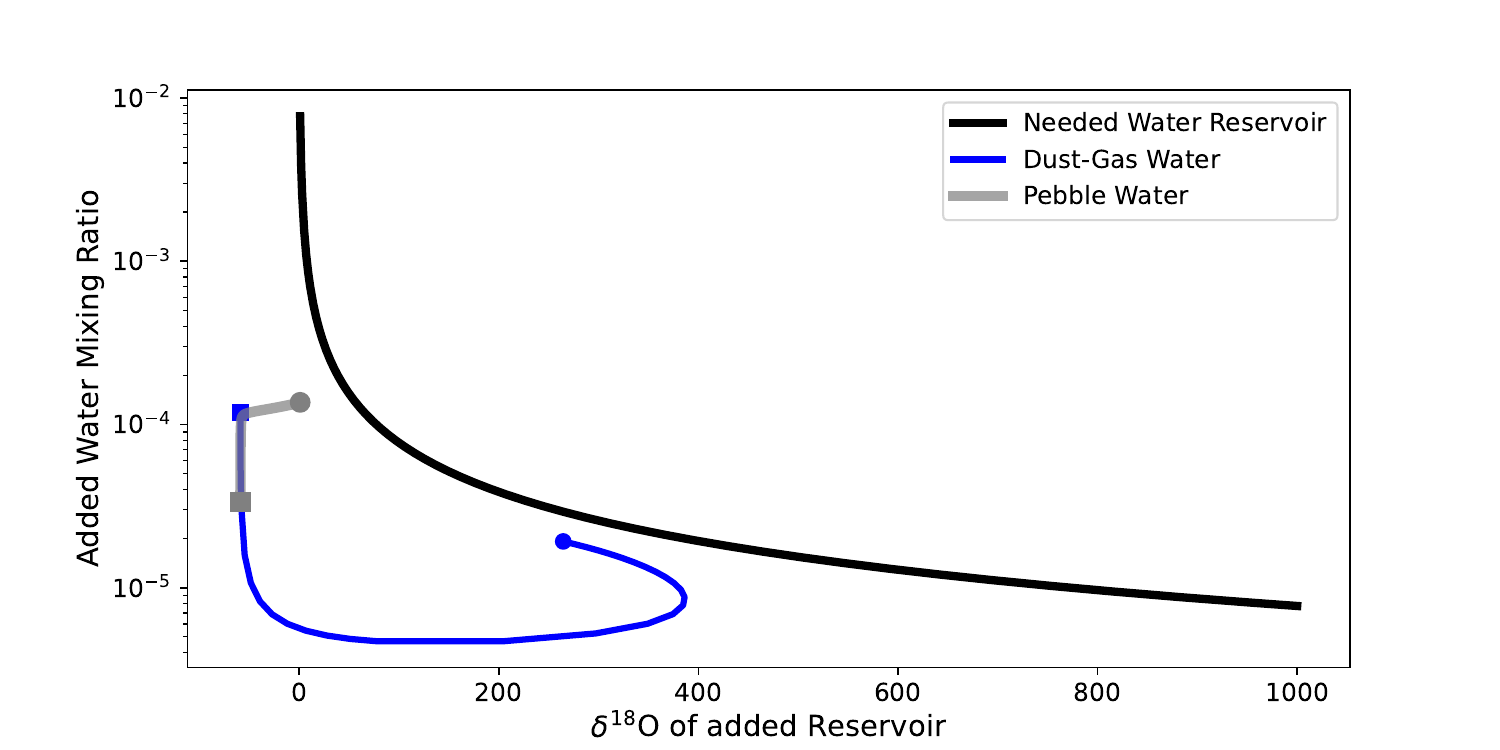}
\includegraphics[width=2.3in]{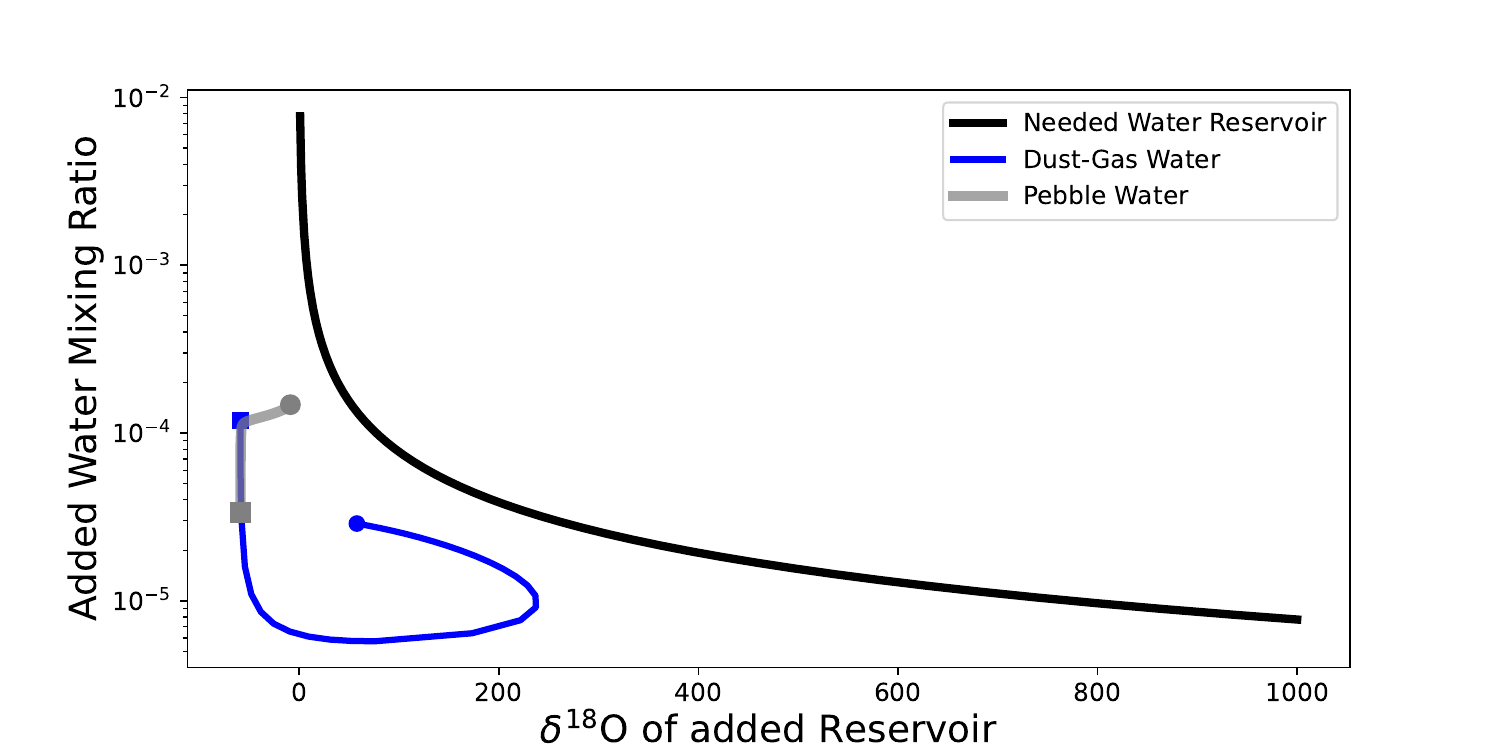} 
    \label{fig:chis} 
    \caption{Comparison of the isotopic evolution of the water reservoirs in the various UV radiation field intensity cases compared to that needed to shift the silicates and oxides of solar composition to terrestrial values, as shown in Figure 4.  The runs are $\chi$=50 (our fiducial run; left), $\chi$=500 (middle) and $\chi$=5000 (right).}
\end{figure}

\subsubsection{Diffusivity} The diffusivity within the disk is highly uncertain, with values of $\alpha$ that characterize the turbulence observationally constrained to be less than 10$^{-3}$ \citep{flaherty2018,dullemond2018}, though higher values have been reported in one disk \citep[IM Lup;][]{paneque-carreno24}.  We thus considered values of $\alpha$=10$^{-4}$, 10$^{-3}$, and 10$^{-2}$, with $\alpha$=10$^{-3}$ corresponding to our fiducial case; the evolutionary pathways for the H$_{2}$O and CO are shown in Figure 9.

Greater levels of diffusion allow for faster delivery of materials from the deep interior of the disk to the photochemically active upper layers.  This results in the CO being mixed through the column on shorter timescales, which allows more to be photodissociated.  This results in a larger water reservoir being produced than in the fiducial case.   Further, the higher diffusivity cases move more CO through the self-shielding region, resulting in greater isotopic fractionation.  Lower levels of diffusion have the opposite effect, leading to less CO being photodissociated and thus less water being produced.

Figure 10 shows how the resulting water reservoirs compare to that which is needed to shift the rocky mineral fractions to the levels seen today. The $\alpha$=10$^{-2}$ case does see the water reservoirs contained in the dust and in the pebbles cross the needed threshold.  Thus a case where vertical mixing timescale is rapid compared to pebble formation timescales could allow for significant incorporation of photochemically-derived products into planetary building blocks.  We return to this point further below when evaluating whether such an outcome could possibly lead to the needed oxygen isotopic evolution for the solar nebula.

 \begin{figure}[!h]
    \includegraphics[width=3.3in]{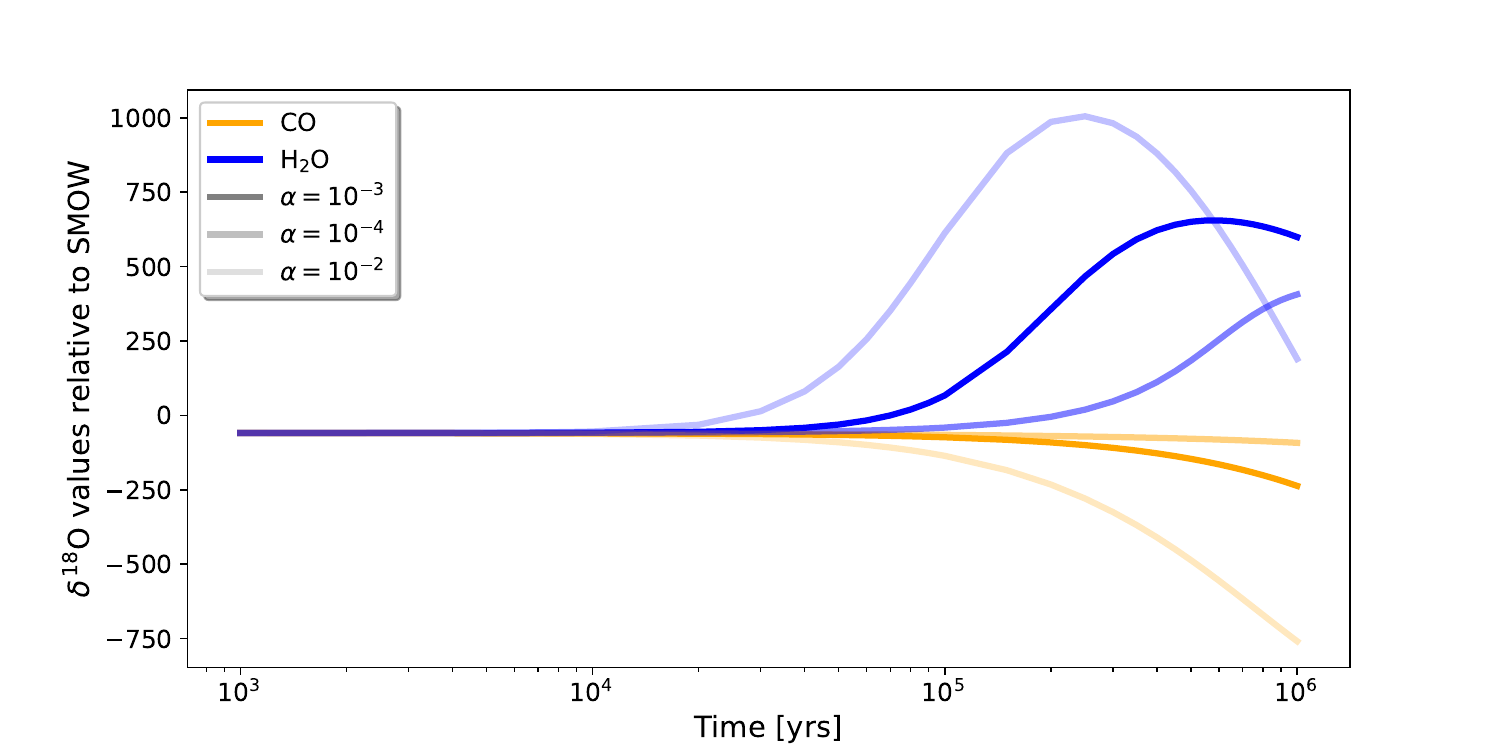} 
    \includegraphics[width=3.3in]{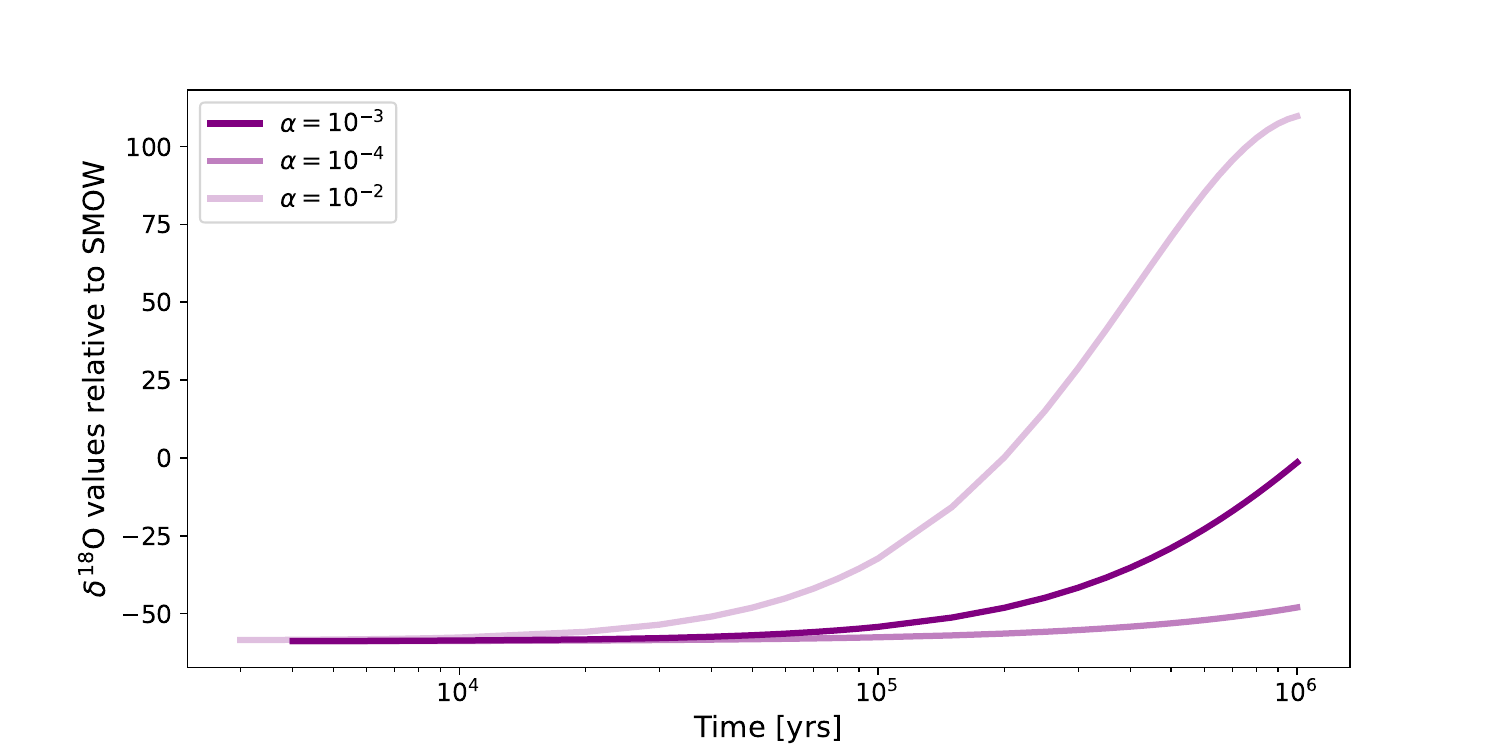} \\
    \label{fig:fig_diff} 
    \caption{Isotopic evolution of the CO and H$_{2}$O reservoirs in the disk (left) and pebbles (right) for various levels of diffusion in the disk.}
\end{figure}

 \begin{figure}[!h]
\includegraphics[width=2.3in]{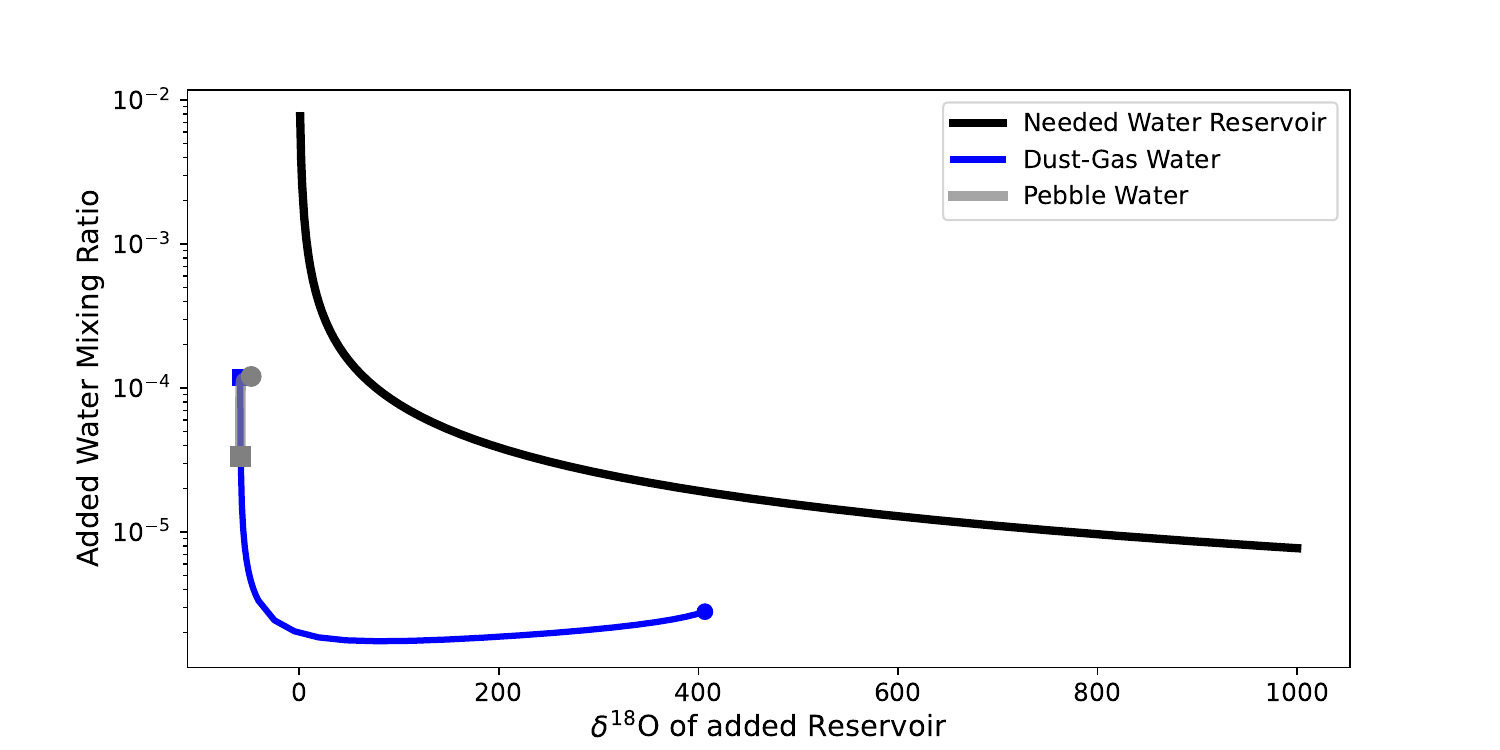}
\includegraphics[width=2.3in]{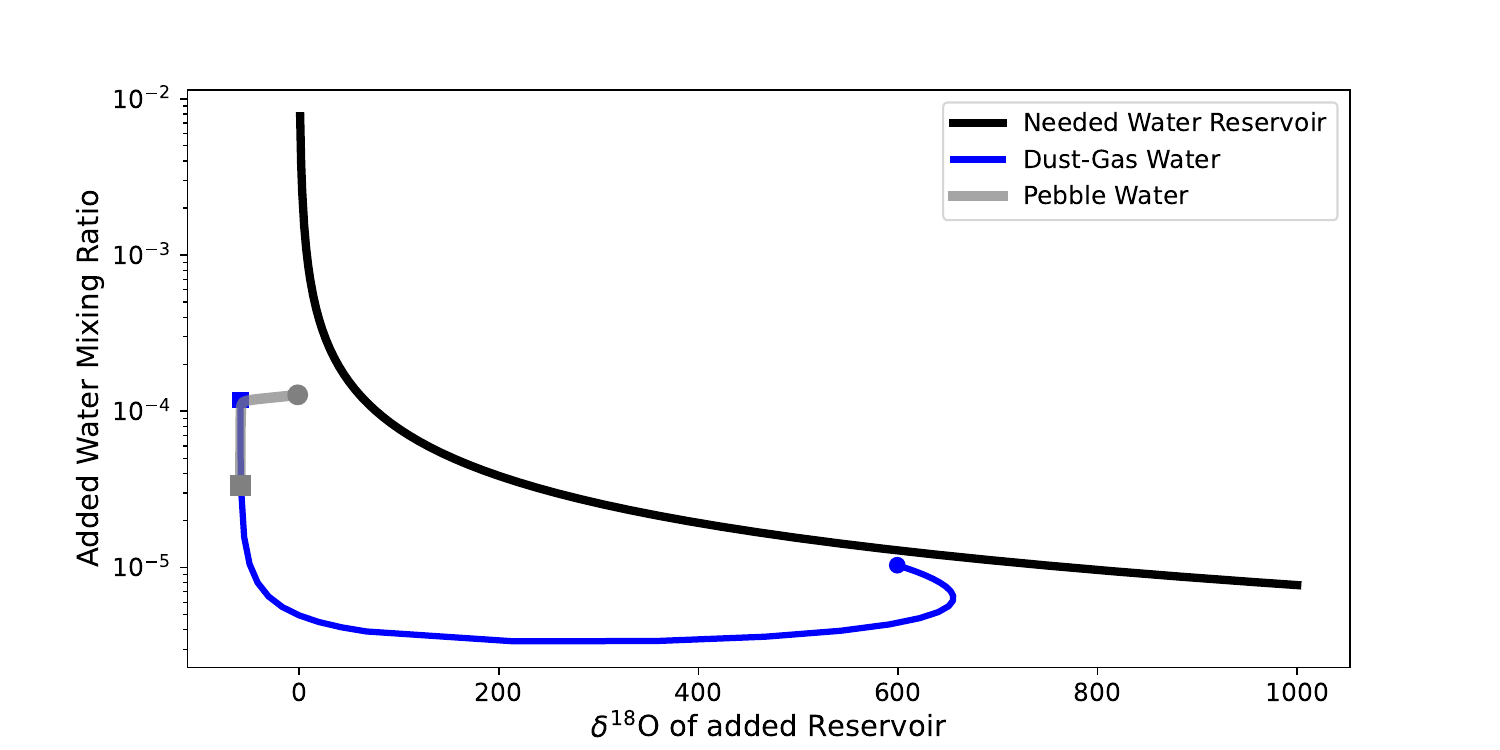}
\includegraphics[width=2.3in]{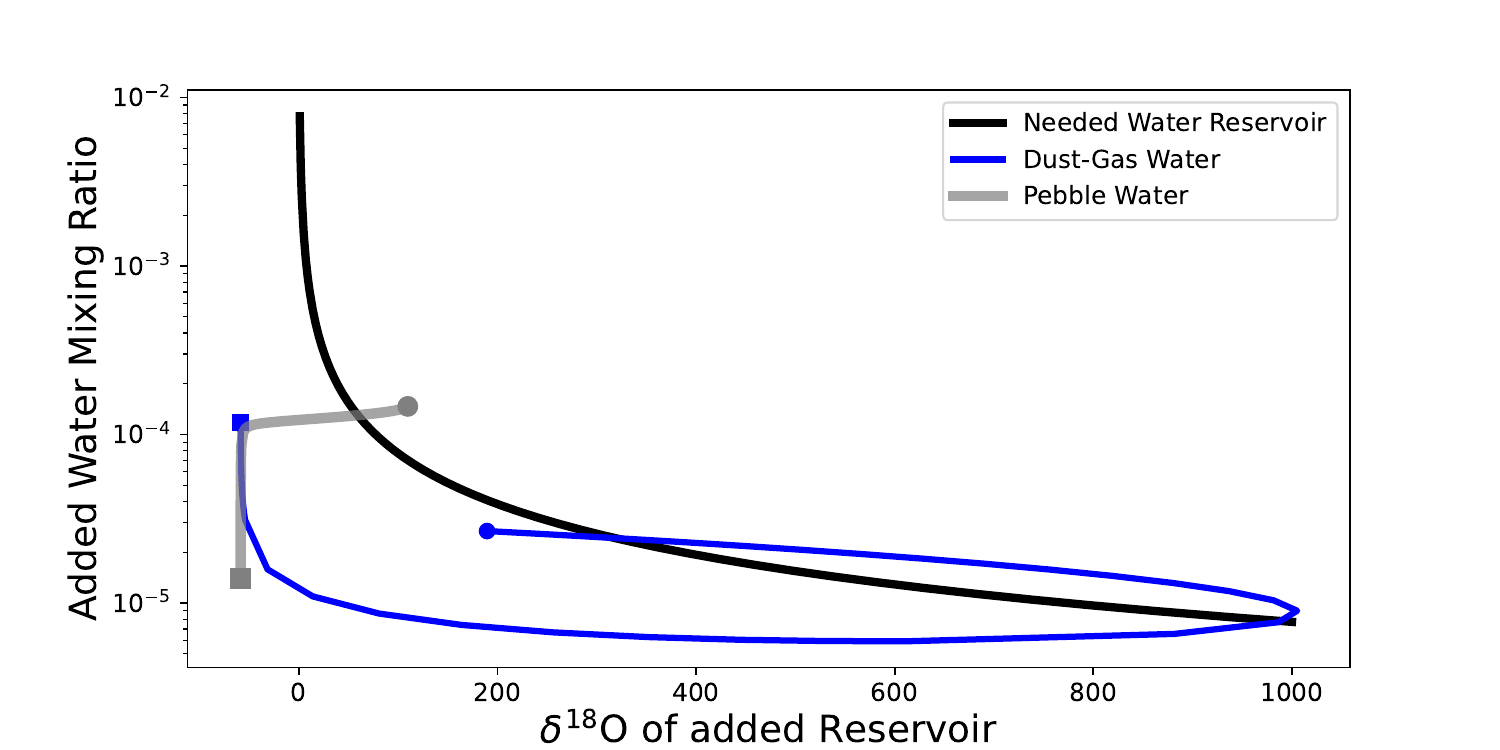} 
    \label{fig:alphas} 
    \caption{Comparison of the isotopic evolution of the water reservoirs in the various diffusivity cases compared to that needed to shift the silicates and oxides of solar composition to terrestrial values, as shown in Figure 4.  The runs are $\alpha$=10$^{-4}$ (left), $\alpha$=10$^{-3}$ (our fiducial run; middle) and $\alpha$=10$^{-2}$ (right).}
\end{figure}

\subsubsection{Distance from Star:}  Figure 11 shows the results of simulations performed at $r$=20, 30, and 40 au from the star in our model disk.  In changing locations, the surface density of the disk changes (greater at 20 au and less at 40 au) as does the diffusive and particle growth timescales (shorter at 20 au and longer at 40 au).  
The overall evolution at 20 and 40 au is similar to that seen in the fiducial case at 30 au.  The more rapid mixing at 20 au leads to slightly more water being produced for the same reasons that increasing the $\alpha$ value does, as described above.  However, in all cases, the pebble reservoir falls short of that needed to reach the oxygen isotopic values observed in the inner disk, though we note the reservoir of water and dust left after 10$^{6}$ years at 20 au is very close to the conditions that would allow the needed shift in oxygen isotopes focused on here.

 \begin{figure}[!h]

    \includegraphics[width=3.3in]{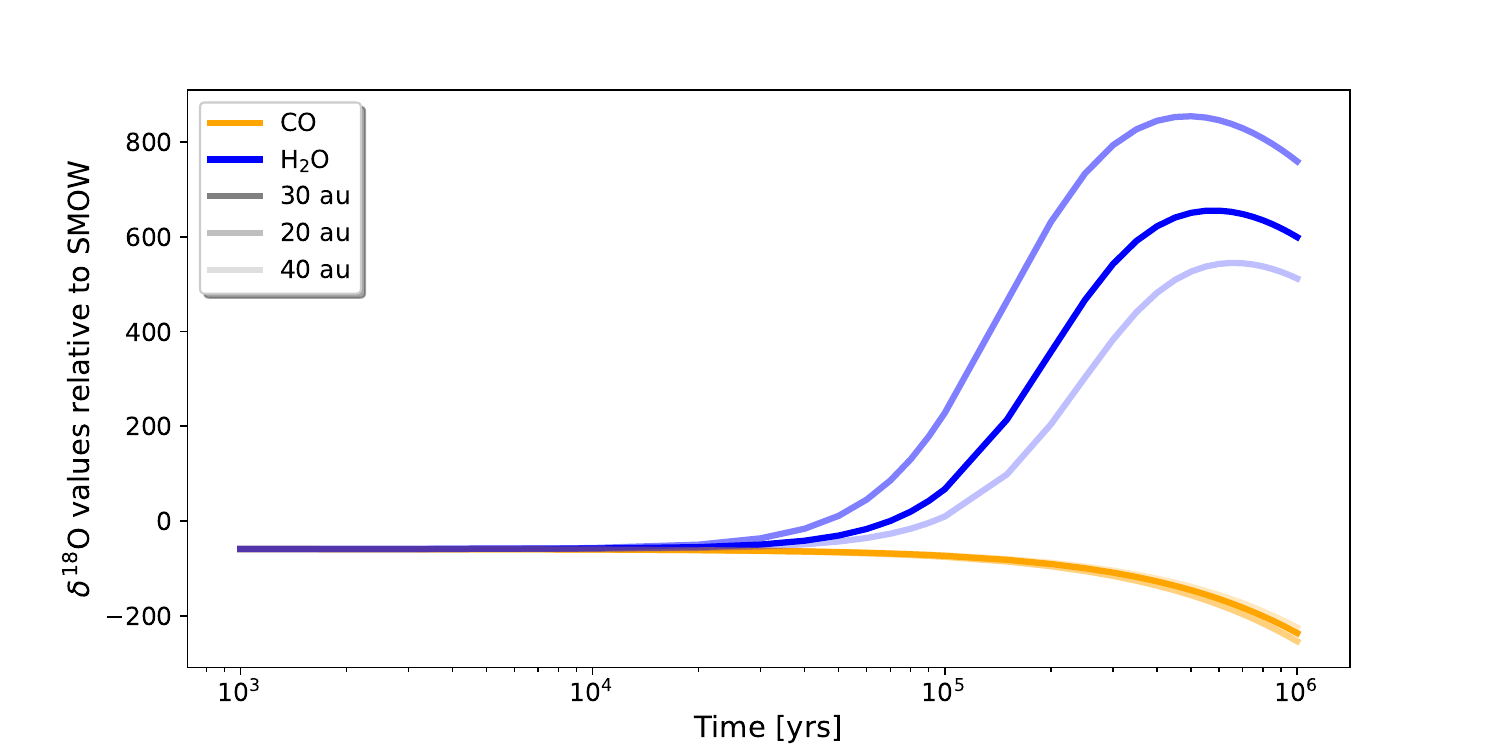} 
    \includegraphics[width=3.3in]{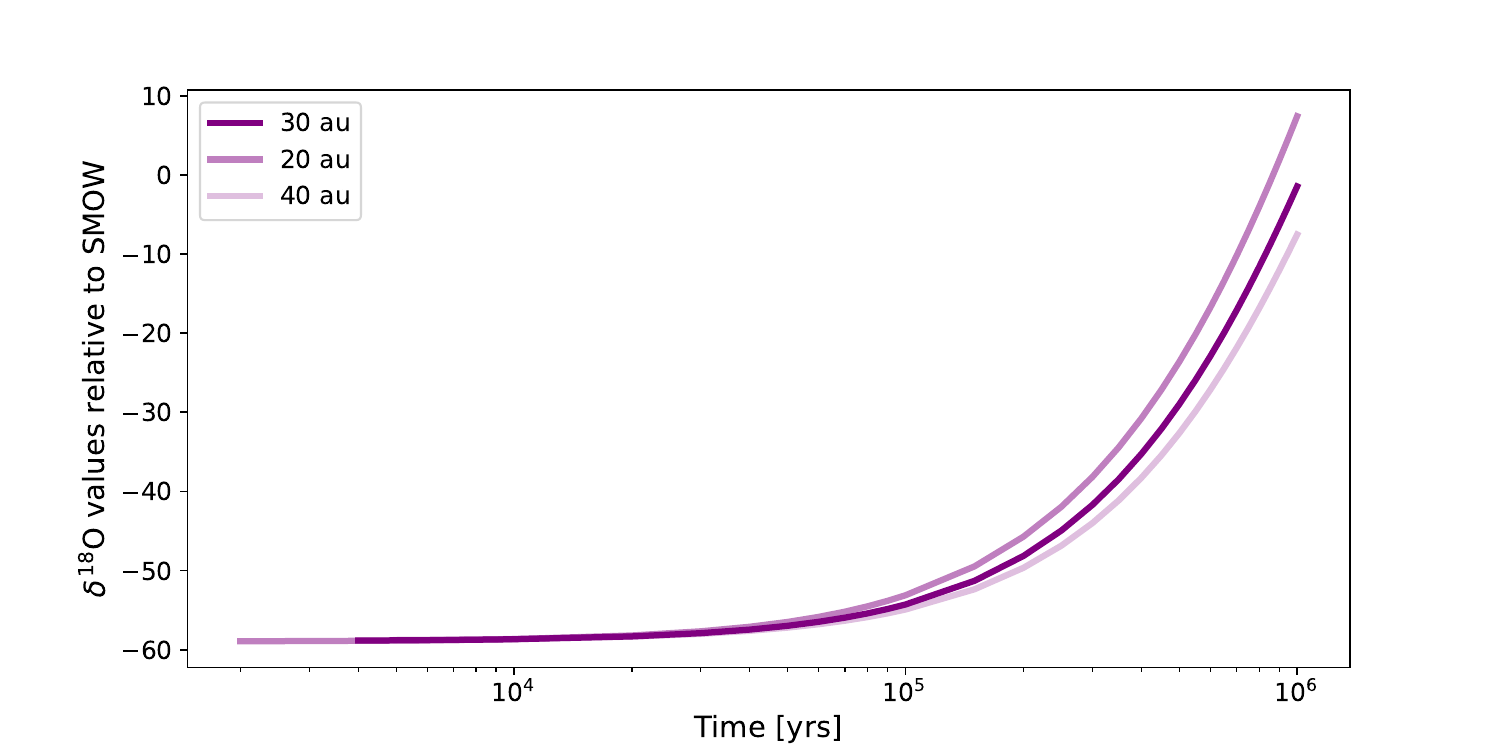} \\
    \label{fig:fig_rs} 
    \caption{Isotopic evolution of the CO and H$_{2}$O reservoirs in the disk (left) and pebbles (right) for various locations in the disk.}
\end{figure}

 \begin{figure}[!h]
 \includegraphics[width=2.3in]{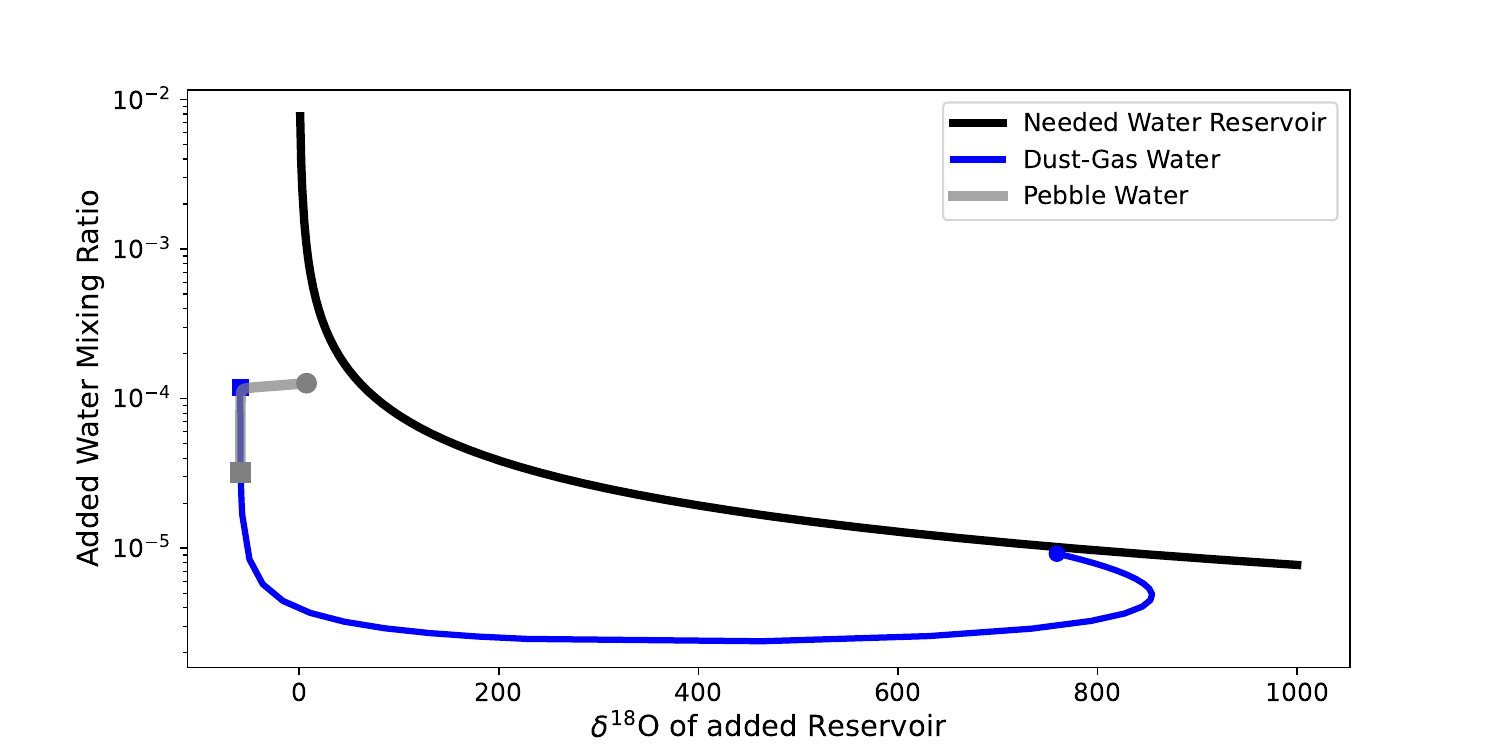}
\includegraphics[width=2.3in]{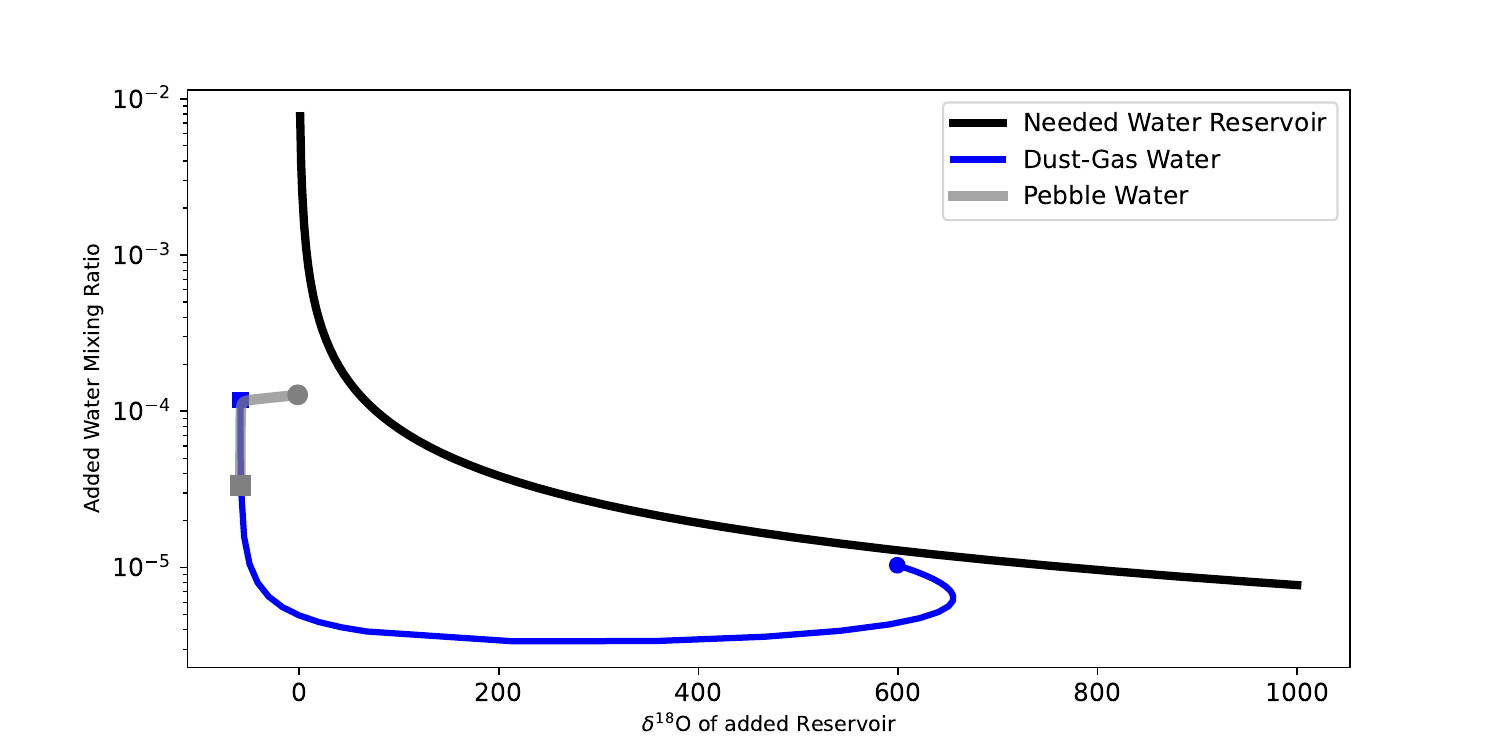}
\includegraphics[width=2.3in]{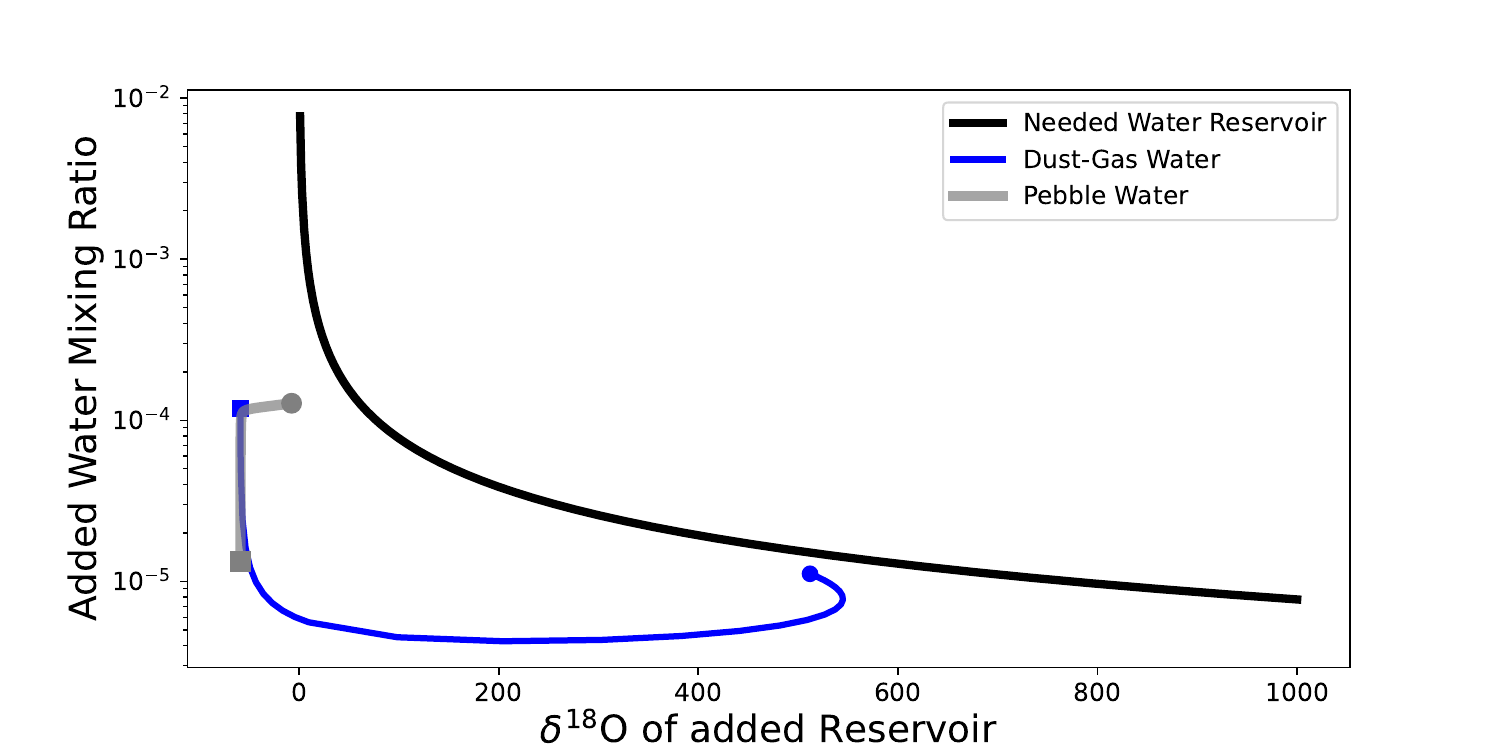} \\
    \label{fig:rs} 
    \caption{Comparison of the isotopic evolution of the water reservoirs in the various locations compared to that needed to shift the silicates and oxides of solar composition to terrestrial values, as shown in Figure 4.  The runs are $r$=20 au (left), $r$=30 au (our fiducial run; middle) and $r$=40 au (right).}
\end{figure}


\subsection{A Closer Look at the Marginal Cases}

While most of the runs outlined above fail to produce a water reservoir that would allow the needed shift in oxygen isotope compositions of silicates and oxides in the outer disk, there are some cases that do satisfy the requirement.  Here we look more closely at these cases to see if we can identify conditions within a disk that would allow the needed water reservoir to develop.

Firstly, the highly diffusive ($\alpha$=10$^{-2}$) case shown in does have ice and pebble reservoirs do contain enough heavy water to meet our constraints.  This is because the high level of diffusion allows more material to be transported through the self-shielding region, enabling a lot of isotopic-selective processing to occur on short timescales. However, it  is worth noting here that while high levels of turbulence have been reported for IM Lup \citep{paneque-carreno24}, most observations constrain $\alpha$ to be no more than $\sim$10$^{-3}$ \citep[e.g.][]{flaherty2018}.  Further, it is generally expected that dust coagulation and the growth of large enough pebbles to decouple from the gas and drift inwards is limited in highly diffusive disks \citep[e.g.][]{birnstiel2012}, and thus pebble growth likely would not proceed as effectively as assumed here.  The high frequency and velocities of collisions expected in such disks would likely slow the rate of fine-dust depletion, making the expected evolution more like those of the slow growth cases shown in Figures 5 and 6.  

Secondly, considering the various processing locations shown in Figures 11 and 12, a trend emerges where the water reservoir grows in abundance and heavy-isotope enrichment as you move to closer distances to the star.  Thus, if as described above, $r$=20 au almost provides the needed water reservoir at the end of the simulation, then one may think that closer distances would be even better locations to produce the needed isotopic evolution.   However, it is important to remember that the simplified chemical network used in this study likely overestimates the rate of water production from the oxygen released from photodissociated CO.  That is, we adopt high values of binding energies for O, OH, and H$_{2}$O compared to literature values; this leads to long residence times of these species on dust grains enabling the hydrogenation reactions that lead to water formation and its fractionation from the CO gas by accumulating into solids.  The binding energies are likely much lower than assumed here, making the photochemically produced water more difficult to form, particularly at elevated temperatures.  To evaluate this, we ran a model at $r$=20 au with the lower binding energies listed in Table 1.  We found that negligible amounts of water formed as the rate of formation was significantly diminished.  This would make isotopic effects even more difficult to develop closer in to the star, where temperatures would be warmer and thus the relevant species would have very short residence times on the dust grains present.  Thus, in all cases presented here, the water produced via photochemistry is likely an overestimate of what would be produced in real protoplanetary disks.

\section{Discussion and Conclusions}

Here we have investigated whether photochemical processing in the solar nebula would be sufficient to produce a reservoir of water sufficiently enriched in the heavy isotopes of oxygen that could then mix with rocky minerals in the inner solar nebula to shift the bulk isotopic ratios from solar to the values observed in meteorites and terrestrial planets.  While isotopically heavy water is produced, we find in all plausible disk conditions and environments, the reservoir that is created and stored in pebbles is not sufficient to match the extent and timing of the record of evolution recorded in these samples.   In general, this is because the isotopically heavy water that is produced is highly diluted by the original inventory of water that was inherited from the molecular cloud.   While preferential sampling of only the later-formed pebbles may be sufficient in some cases to get close to that needed to match what is seen in planetary and asteroidal samples, the later times would be inconsistent with the chronology of evolution recorded by chondritic meteorites.  Further, it would require some explanation for why we see no record of the very large inventory of the early-formed pebbles and the solar-like water they contained. 

As a result, we rule out photoprocessing of CO in the solar nebula as being primarily responsible for the enrichment of rocky minerals in the heavy isotopes of oxygen compared to the Sun.  If photochemical processing is required to drive this evolution via creation of a reservoir enriched in heavy isotopes of oxygen, it must have occurred within the natal molecular cloud \citep{yurimoto2004, lee2008}, with the carriers surviving their incorporation into the solar nebula.  It is possible that photochemical processing in the disk did contribute to some of the lower magnitude variations seen across solar system rocks.

The results provided here are not limited to just oxygen isotope evolution, but can be generalized to all photochemical products.  For instance, self shielding in the solar nebula has been suggested to contribute to nitrogen isotopic anomalies \citep{Chakraborty2014} and photoprocessing also could contribute to deuterium anomalies in water and other molecules \citep{Cleeves2014}.  However, the necessary photochemical reactions are limited to the optically thin, upper layers of the disk.  Generally, this region, by mass, makes up a just a small fraction of the disk, and thus the photochemical products are diluted by what was already present in the disk. While the penetration depth of the photons may grow as dust grains grow, their growth predates this onset of greater photochemical processing, making them unable to incorporate the products of those reactions.  

Further, these results have implications for our interpretations of observations of other protoplanetary disks.  While the impacts of photochemistry are readily seen in the gas-phase abundances of molecules in the disks \citep{bergner2021,oberg2021}, this chemistry is likely not significantly impacting the compositions of the solids that ultimately form planets.  In fact, we find here that there can be a significant compositional difference in the molecules that remain in the gas and mantles of fine-dust in the disk, which is what we typically observe, compared to that which is contained in the larger aggregates which serve as the building blocks of the planets.  Thus, those molecules that are observed in evolved protoplanetary disks are not necessarily present on the solids that would be accreted by growing planets.

\begin{acknowledgments}
FJC and EB acknowledge funding by NASA grants 80NSSC20K0333 and 80NSSC20K0259.  EVC acknowledges support from NASA FINESST grant 80NSSC23K1380.  JB acknowledges support from NASA's Hubble Fellowship Program.  The authors are greatful for the constructive and insightful comments provided by the referee which improved this paper.
\end{acknowledgments}

\software{
\texttt{CANDY} \citep{vanclepper2022}
\texttt{astrochem} \citep{maret_bergin2015}
\texttt{matplotlib} \citep{hunter_matplotlib_2007}
\texttt{numpy} \citep{harris_array_2020}
          }


\bibliographystyle{aasjournalv7}



\end{document}